\documentclass[longauth]{aa}
\usepackage[varg]{txfonts}
\usepackage[colorlinks,urlcolor=blue,citecolor=blue,linkcolor=blue]{hyperref} 
\usepackage{longtable,verbatim,amsmath,natbib,graphicx}

\def\oldlabel#1{{}}
\def\bfref{}

\title{The LOFAR Pilot Surveys for Pulsars and Fast Radio
  Transients\thanks{\url{http://www.astron.nl/pulsars/lofar/surveys/lotas/}}}
\author{
  Thijs Coenen    \inst{\ref{uva} \and \ref{astron}} \and
  Joeri van Leeuwen     \inst{\ref{astron}
\thanks{e-mail: \href{mailto:leeuwen@astron.nl}{leeuwen@astron.nl}}
 \and \ref{uva}} \and
  Jason W.~T.~Hessels   \inst{\ref{astron} \and \ref{uva}} \and
  Ben~W.~Stappers       \inst{\ref{jod}}   \and 
  Vladislav~I.~Kondratiev       \inst{\ref{astron} \and \ref{leb}} \and \\
      A.~Alexov  \inst{ \ref{uva} \and \ref{stsci}} \and
      R.~P.~Breton  \inst{ \ref{soton}}    \and
      A.~Bilous  \inst{ \ref{nijmegen}} \and
      S.~Cooper  \inst{\ref{jod}}       \and
      H.~Falcke  \inst{ \ref{nijmegen} \and \ref{astron}} \and
      R.~A.~Fallows\inst{ \ref{astron}} \and
      V.~Gajjar \inst{\ref{ncra}},
       \inst{\ref{jod}},
      J.-M.~Grie\ss{}meier\inst{  \ref{cnrs} \and \ref{nancay}} \and
      T.~E.~Hassall\inst{ \ref{soton}} \and
      A.~Karastergiou\inst{ \ref{ox}} \and
      E.~F.~Keane\inst{ \ref{swin} \and \ref{caastro}} \and
      M.~Kramer\inst{ \ref{mpifr} \and \ref{jod}} \and
      M.~Kuniyoshi\inst{ \ref{mpifr}} \and
      A.~Noutsos\inst{ \ref{mpifr}} \and
      S.~Os{\l}owski  \inst{ \ref{mpifr} \and \ref{biel}} \and
      M.~Pilia        \inst{ \ref{astron}} \and
      M.~Serylak\inst{ \ref{ox}} \and
      C.~Schrijvers\inst{\ref{surfsara}}, 
      C.~Sobey     \inst{ \ref{astron}}  \and
      S.~ter~Veen  \inst{ \ref{nijmegen}}\and
      J.~Verbiest  \inst{ \ref{biel}} \and
      P.~Weltevrede    \inst{ \ref{jod}} \and
      S.~Wijnholds \inst{\ref{astron}} \and
      K.~Zagkouris     \inst{ \ref{ox}}
\and A.S.~van Amesfoort   \inst{\ref{astron}}
\and J.~Anderson          \inst{\ref{hzp} \and \ref{aip}}
\and A.~Asgekar           \inst{\ref{astron} \and \ref{shell}}
\and I.~M.~Avruch         \inst{\ref{sron} \and \ref{kapteyn}}
\and M.~E.~Bell           \inst{\ref{csiro}}
\and M.~J.~Bentum         \inst{\ref{astron} \and \ref{twente}}
\and G.~Bernardi          \inst{\ref{cfa}}  
\and P.~Best              \inst{\ref{roe}}
\and A.~Bonafede          \inst{\ref{hamburg}}
\and F.~Breitling         \inst{\ref{aip}}
\and J.~Broderick         \inst{\ref{soton}}
\and M.~Br\"uggen         \inst{\ref{hamburg}}
\and H.~R.~Butcher        \inst{\ref{anu}}
\and B.~Ciardi            \inst{\ref{mpifa}}
\and A.~Corstanje         \inst{\ref{nijmegen}}
\and A.~Deller            \inst{\ref{astron}}
\and S.~Duscha            \inst{\ref{astron}}
\and J.~Eisl\"offel       \inst{\ref{tls}}
\and R.~Fender            \inst{\ref{ox}}
\and C.~Ferrari           \inst{\ref{nice}}
\and W.~Frieswijk         \inst{\ref{astron}}
\and M.~A.~Garrett        \inst{\ref{astron} \and \ref{leiden}}
\and F.~de Gasperin       \inst{\ref{hamburg}}
\and E.~de Geus           \inst{\ref{astron} \and \ref{smartervision}}
\and A.~W.~Gunst          \inst{\ref{astron}}
\and J.~P.~Hamaker        \inst{\ref{astron}}
\and G.~Heald             \inst{\ref{astron}}
\and M.~Hoeft             \inst{\ref{tls}}
\and A.~van der Horst     \inst{\ref{uva}}
\and E.~Juette            \inst{\ref{raiub}}
\and G.~Kuper             \inst{\ref{astron}}
\and C.~Law               \inst{\ref{berkley} \and \ref{uva}}
\and G.~Mann              \inst{\ref{aip}}
\and R. McFadden          \inst{\ref{astron}}
\and D.~McKay-Bukowski    \inst{\ref{sodankyla} \and \ref{stfc}}
\and J.~P.~McKean         \inst{\ref{astron} \and \ref{kapteyn}}
\and H.~Munk              \inst{\ref{astron}}
\and E.~Orru              \inst{\ref{astron}}
\and H.~Paas              \inst{\ref{groningen}}
\and M.~Pandey-Pommier    \inst{\ref{lyon}}
\and A.~G.~Polatidis      \inst{\ref{astron}}
\and W.~Reich             \inst{\ref{mpifr}}
\and A.~Renting           \inst{\ref{astron}}
\and H.~R\"ottgering      \inst{\ref{leiden}}
\and A.~ Rowlinson        \inst{\ref{uva}}
\and A.~M.~M.~Scaife      \inst{\ref{soton}}
\and D.~Schwarz           \inst{\ref{biel}}
\and J.~Sluman            \inst{\ref{astron}}
\and O.~Smirnov           \inst{\ref{crat} \and \ref{skasa}}
\and J.~Swinbank          \inst{\ref{uva}}
\and M.~Tagger            \inst{\ref{cnrs}}
\and Y.~Tang              \inst{\ref{astron}}
\and C.~Tasse             \inst{\ref{meudon}}
\and S.~Thoudam           \inst{\ref{nijmegen}}
\and C.~Toribio           \inst{\ref{astron}}
\and R.~Vermeulen         \inst{\ref{astron}}
\and C.~Vocks             \inst{\ref{aip}}
\and R.~J.~van Weeren     \inst{\ref{cfa}}
\and O.~Wucknitz          \inst{\ref{mpifr}}
\and P.~Zarka             \inst{\ref{meudon}}
\and A.~Zensus            \inst{\ref{mpifr}}
}

\institute{
 Anton Pannekoek Institute for Astronomy, University of Amsterdam, Science Park 904,
             1098 XH Amsterdam, The Netherlands\label{uva}
  \and
ASTRON, the Netherlands Institute for Radio Astronomy, Postbus 2, 7990
AA, Dwingeloo, The Netherlands\label{astron}
  \and 
Jodrell Bank Center for Astrophysics, School of Physics and Astronomy,
University of Manchester, Manchester M13 9PL, UK\label{jod}
  \and
Astro Space Center of the Lebedev Physical Institute, Profsoyuznaya str. 84/32, Moscow 117997, Russia\label{leb}
  \and
Space Telescope Science Institute, 3700 San Martin Drive, Baltimore, MD 21218, USA \label{stsci} 
  \and
School of Physics and Astronomy, University of Southampton, Southampton, SO17 1BJ, UK \label{soton}
  \and
Department of Astrophysics/IMAPP, Radboud University Nijmegen, P.O. Box 9010, 6500 GL Nijmegen, The Netherlands \label{nijmegen} 
  \and
National Centre for Radio Astrophysics, Post Bag 3, Ganeshkhind, Pune 411 007, India\label{ncra}   \and
Laboratoire de Physique et Chimie de l'Environnement et de l'Espace, LPC2E UMR 7328 CNRS, 45071 Orl\'{e}ans, France \label{cnrs} 
  \and
Station de Radioastronomie de Nan\c{c}ay, Observatoire de Paris, CNRS/INSU, 18330 Nan\c{c}ay, France \label{nancay}   \and
Astrophysics, University of Oxford, Denys Wilkinson Building, Keble Road, Oxford OX1 3RH \label{ox} 
  \and
Centre for Astrophysics and Supercomputing, Swinburne University of Technology, Mail H30, PO Box 218, VIC 3122, Australia \label{swin}
  \and
ARC Centre of Excellence for All-sky Astrophysics (CAASTRO).\label{caastro}
  \and
Max-Planck-Institut f\"{u}r Radioastronomie, Auf dem H\"ugel 69, 53121 Bonn, Germany \label{mpifr} 
  \and
Fakult{\"a}t f{\"u}r Physik, Universit{\"a}t Bielefeld, Postfach 100131, D-33501 Bielefeld, Germany \label{biel} 
  \and
SURFsara, PO Box 94613, NL-1090 GP Amsterdam, the Netherlands\label{surfsara} 
  \and
Helmholtz-Zentrum Potsdam, DeutschesGeoForschungsZentrum GFZ, Department 1: Geodesy and Remote Sensing, Telegrafenberg, A17, 14473 Potsdam, Germany \label{hzp}
  \and
Leibniz-Institut f\"{u}r Astrophysik Potsdam (AIP), An der Sternwarte 16, 14482 Potsdam, Germany \label{aip}
  \and
Shell Technology Center, Bangalore, India \label{shell}
  \and
SRON Netherlands Insitute for Space Research, PO Box 800, 9700 AV Groningen, The Netherlands \label{sron}
  \and
Kapteyn Astronomical Institute, PO Box 800, 9700 AV Groningen, The Netherlands \label{kapteyn}
  \and
CSIRO Australia Telescope National Facility, PO Box 76, Epping NSW 1710, Australia \label{csiro}
  \and
University of Twente, The Netherlands \label{twente}
  \and
Harvard-Smithsonian Center for Astrophysics, 60 Garden Street, Cambridge, MA 02138, USA \label{cfa}
  \and
Institute for Astronomy, University of Edinburgh, Royal Observatory of Edinburgh, Blackford Hill, Edinburgh EH9 3HJ, UK \label{roe}
  \and
University of Hamburg, Gojenbergsweg 112, 21029 Hamburg, Germany \label{hamburg}
  \and
Research School of Astronomy and Astrophysics, Australian National University, Mt Stromlo Obs., via Cotter Road, Weston, A.C.T. 2611, Australia \label{anu}
  \and
Max Planck Institute for Astrophysics, Karl Schwarzschild Str. 1, 85741 Garching, Germany \label{mpifa}
  \and
Th\"{u}ringer Landessternwarte, Sternwarte 5, D-07778 Tautenburg, Germany \label{tls}
  \and
Laboratoire Lagrange, UMR7293, Universit\`{e} de Nice Sophia-Antipolis, CNRS, Observatoire de la C\'{o}te d'Azur, 06300 Nice, France \label{nice}
  \and
Leiden Observatory, Leiden University, PO Box 9513, 2300 RA Leiden, The Netherlands \label{leiden}
  \and
SmarterVision BV, Oostersingel 5, 9401 JX Assen \label{smartervision}
  \and
Astronomisches Institut der Ruhr-Universit\"{a}t Bochum, Universitaetsstrasse 150, 44780 Bochum, Germany \label{raiub}
  \and
Radio Astronomy Lab, UC Berkeley, CA, USA \label{berkley}
  \and
Sodankyl\"{a} Geophysical Observatory, University of Oulu, T\"{a}htel\"{a}ntie 62, 99600 Sodankyl\"{a}, Finland \label{sodankyla}
  \and
STFC Rutherford Appleton Laboratory,  Harwell Science and Innovation Campus,  Didcot  OX11 0QX, UK \label{stfc}
  \and
Center for Information Technology (CIT), University of Groningen, The Netherlands \label{groningen}
  \and
Centre de Recherche Astrophysique de Lyon, Observatoire de Lyon, 9 av Charles Andr\'{e}, 69561 Saint Genis Laval Cedex, France \label{lyon}
  \and
Department of Physics and Elelctronics, Rhodes University, PO Box 94, Grahamstown 6140, South Africa \label{crat}
  \and
SKA South Africa, 3rd Floor, The Park, Park Road, Pinelands, 7405, South Africa \label{skasa}
  \and
LESIA, UMR CNRS 8109, Observatoire de Paris, 92195   Meudon, France \label{meudon}
}

\begin{document}

\abstract{
We have conducted two pilot surveys for radio pulsars and fast
transients with the Low-Frequency Array 
(LOFAR) around 140\,MHz and here report on the first low-frequency fast-radio burst
limit and the discovery of two new pulsars.
The first survey, the LOFAR Pilot Pulsar Survey (LPPS), observed a
large fraction of the northern sky, $\sim$$1.4\times10^4$ \,deg$^2$,
with 1-hr dwell times. Each observation covered $\sim$75\,deg$^2$
using 7 independent fields formed by incoherently summing the
high-band antenna fields. The second pilot survey, the LOFAR
Tied-Array Survey (LOTAS), spanned $\sim$600\,deg$^2$, with roughly a
5-fold increase in sensitivity compared with LPPS. Using a coherent
sum of the 6 LOFAR ``Superterp'' stations, we formed 19 tied-array
beams, together covering 4\,deg$^2$ per pointing.  
 From LPPS we
derive a limit on the occurrence, at 142\,MHz, of
dispersed radio bursts of $< 150$\,day$^{-1}$\,sky$^{-1}$, 
for bursts brighter than $S > 107$\,Jy for the narrowest
 searched burst duration of 0.66\,ms.
In LPPS, we re-detected 65 previously known pulsars.  LOTAS discovered
two pulsars, the first with LOFAR or any digital aperture array. LOTAS
also re-detected 27 previously known pulsars.  These pilot studies
show that LOFAR can efficiently carry out all-sky surveys for pulsars
and fast transients, and they set the stage for further surveying
efforts using LOFAR and the planned low-frequency component of the
Square Kilometer Array.
}

\maketitle

\section{Introduction}\label{lpps-section-introduction}

The Low-Frequency Array \citep[LOFAR;][]{hwg+13}, with its high
sensitivity and flexible observing configurations, is set to open the
lowest radio frequencies to efficient pulsar surveys. Its operating
frequency of 10$-$240\,MHz means a return to the long-wavelength range
at which pulsars were originally discovered.
The first pulsars were detected  at $81.5\,\mathrm{MHz}$, with the
Cambridge Inter Planetary Scattering array
\citep{1968Natur.217..709H}.  Since then, however, most pulsar surveys
have avoided low radio frequencies ($<300\;\mathrm{MHz}$) because of a
number of effects that scale strongly with decreasing frequency and
severely affect pulsar detectability \citep{2011A&A...530A..80S}.
First, at low frequencies, dispersion correction requires many more,
narrower frequency channels \citep{2005hpa..book.....L}, and searches
need a much finer grid of trial dispersion measures (DMs).  Second,
multi-path propagation caused by interstellar scattering broadens the
intrinsically short duration pulses, scaling with frequency $\nu$ as $\nu^{-3.9}$
\citep{2004ApJ...605..759B}.  Lastly, the sky background temperature
$T_\mathrm{sky}$ increases at low frequencies as $\nu^{-2.6}$
\citep{1987MNRAS.225..307L}.

These drawbacks are partially compensated for by the steep pulsar
spectral indices, $S \propto \nu^{-1.8}$ on average
(\citealt{2000A&AS..147..195M}; interpreted further in
\citealt{2013MNRAS.431.1352B}), with outliers of $S \propto \nu^{-2}$
- $\nu^{-4}$ \citep{hsh+14}.  Furthermore, as many pulsar spectra turn
over toward lower frequencies, flux densities typically peak in the 100-200\,MHz
band \citep{2000ARep...44..436M,hsh+14}.  Finally, for some pulsars
only the wider, low-frequency beam may cross Earth
\citep{2011A&A...530A..80S}.  Overall these make a compelling
low-frequency search case.

A number of low-frequency surveys have
recently been performed between 16 and 400\,MHz:
The Ukrainian UTR-2 radio telescope carried out a pulsar census in the
$16.5-33.0\,\mathrm{MHz}$ band \citep{2013MNRAS.431.3624Z}.
A 34.5-MHz pulsar survey undertaken with the Gauribidanur telescope
resulted in the first (potential) radio
detection of {\it Fermi} pulsar J1732$-$3131
\citep{2012MNRAS.425....2M} --- a source that might only be detectable
below 50\,MHz.
The Cambridge array performed a second pulsar survey at 81.5\,MHz
but did not discover any new pulsars \citep{1998ApJ...509..785S}.
The ongoing Arecibo drift survey for pulsars at 327\,MHz discovered 24
new pulsars \citep{2013ApJ...775...51D}.
A survey of the Cygnus region using the Westerbork Synthesis Radio
Telescope (WSRT) at 328\,MHz led to the discovery of 3 new pulsars
\citep{2009A&A...498..223J}, also demonstrating the use of a
dish-based radio interferometer for pulsar searching.
Perhaps most importantly, several $300 - 400$\,MHz surveys have recently been
conducted using the Green Bank Telescope (GBT).  Through their high time
and frequency resolution, these surveys have
discovered over 100 normal and millisecond pulsars so far
\citep[][]{2008AIPC..983..613H,bls+13,lbr+13,slr+14}.

Low-frequency surveys can also achieve large instantaneous sky
coverage and sensitivity. That is important, as over the last decade it has become
increasingly apparent that the various sub-types of radio-emitting
neutron stars show a wide range of activity -- from the classical,
steady pulsars to the sporadic pulses of the rotating radio transients
\citep[RRATs;][]{2006Natur.439..817M}, and the off-on intermittent
pulsars \citep{2006Sci...312..549K}. Other cases of transient
millisecond radio pulsars
\citep[][]{2009Sci...324.1411A,2013ATel.5069....1P,2014MNRAS.441.1825B,2014ApJ...790...39S}
and radio 
magnetars \citep{2006Natur.442..892C,2013Natur.501..391E} also give strong motivation for
pulsar surveys that permit large on-sky time and repeated observations
of the same survey area.  Furthermore, the recent discovery of the
fast radio bursts \citep[FRBs, also known as ``Lorimer
  Bursts'';][]{2007Sci...318..777L,2012MNRAS.425L..71K,2013Sci...341...53T,2014arXiv1404.2934S,2014arXiv1405.5945P}
provides even more impetus for wide-field radio surveys with
sub-millisecond time resolution.


LOFAR  operates in two
observing bands by using two different types of antennas: the LOFAR
low-band antennas (LBA; 10$-$90\,MHz) and the LOFAR high-band antennas
(HBA; 110$-$240\,MHz), which together cover the lowest 4 octaves of
the radio window observable from Earth. It has a dense
central core in the north of the Netherlands. 
Core HBA stations consist of
two identically sized ``sub-stations'' \cite[Fig.~4
  in][]{hwg+13}. LOFAR uses digital electronics 
and high-speed fiber networks to create a low-frequency
interferometric array that is more versatile and scientifically
capable than its predecessors.  In particular, through the use of
multi-beaming techniques LOFAR can achieve very large (tens of square
degrees) fields-of-view, which are ideal for performing efficient
pulsar and fast transient surveys.  The wide fractional bandwidth and
capability to store and process hundreds of terabytes of
high-resolution data make it an even more powerful instrument. A
general overview of LOFAR's pulsar observing capabilities is provided
in \cite{2011A&A...530A..80S}, while LOFAR pulsar-survey strategies
are outlined and simulated in \citet{ls10}.

\begin{figure}[t]
\centering
\includegraphics[width=\columnwidth]{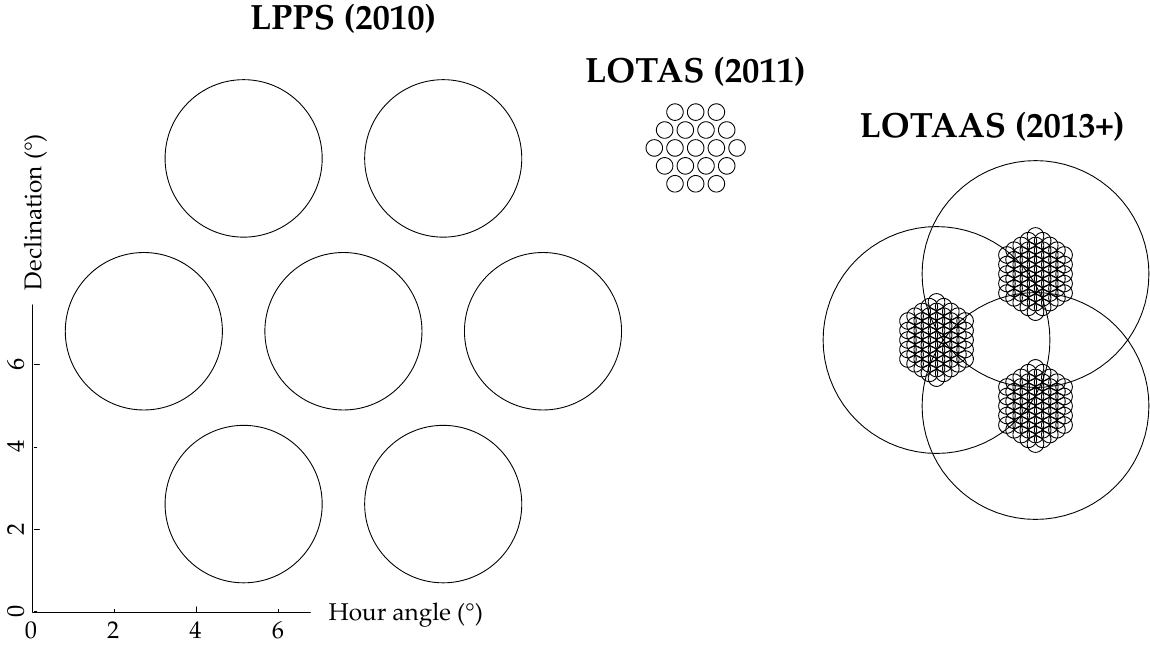}
\caption{The single-pointing footprints of LPPS (left) and LOTAS
  (middle).  Circles denote the half-power beam widths at the respective central frequencies (Table \ref{table-obs}).  For
  comparison (right) is the single-pointing footprint of the ongoing
  LOFAR Tied-array All-sky Survey {LOTAAS} (Section
  \ref{overall-section-conclusion}).  The LOTAAS pointings use both
  incoherent (large circles) and coherent (clusters of small circles)
  beams.}
\label{fig-lotas-pointing-comparison}
\end{figure}

During the $2008 - 2012$ commissioning of LOFAR, we performed two
pilot pulsar surveys. Efficient pulsar surveys ideally combine high
sensitivity, a large  total field-of-view (FoV)
and a high angular resolution per FoV element.  LOFAR's {beam-formed} modes
\citep{2011A&A...530A..80S} can form up to hundreds of beams
simultaneously \citep{Romein:10}, and thus provide such a wide view
\mbox{($>$10\;deg$^2$)} as well as the ability to constrain positions to a
few arcminutes.
The first survey, employing \emph{incoherent}
beam-forming,
is called the LOFAR Pilot Pulsar Survey (LPPS;
Fig. \ref{fig-lotas-pointing-comparison}, left panel), while the
second survey, the LOFAR Tied-Array Survey (LOTAS;
Fig. \ref{fig-lotas-pointing-comparison}, middle panel) employed
\emph{coherent} or {tied-array} beam forming. 
Incoherent beam-forming \citep{bac99}
provides lower raw sensitivity, with pulsar signal-to-noise (S/N) increasing only as the square root of the
number of stations being added.
But it does allow one to observe bright,
 rare events because of its large FoV, clear from
Fig. \ref{fig-lotas-pointing-comparison}. 
Coherent beam-forming, in contrast,
permits maximum instantaneous sensitivity, scaling linearly with the number
of stations summed. The resulting tied-array beams have limited FoV,
however, scaling as the
inverse of the maximum distance between stations \citep{2011A&A...530A..80S}.

\defcitealias{coen13}{C13}
In this paper, we present the setup of these two surveys
(Sect. \ref{sect:obs} and \ref{sect:analyse}). In Section
\ref{sec:results} we then  
report on the new and known pulsars that were detected and
present limits for the rate of  fast radio transients in general.
Section \ref{overall-section-conclusion} contains overall conclusions,
and looks forward to currently ongoing and future work. 
Profiles and parameters for the detected and discovered pulsars can be
found in Appendices \ref{sec:positions:lpps} and
\ref{sec:positions:lotas}. These sections are all described in greater
detail in \citet[][henceforth \citetalias{coen13}]{coen13}, 
and throughout this paper we will refer the interested reader to
specific sections of \citetalias{coen13}.

\section{Observations} \label{sect:obs}
In LPPS we  \emph{incoherently} added as many LOFAR stations as were
available; For LOTAS we only \emph{coherently} combined stations from the
300-m-wide LOFAR ``Superterp'', which has a much higher filling factor
\citep{bac99,ls10} than the rest of the array (Table \ref{table-obs}).

\begin{table*}[bt]
\centering
\begin{tabular}{l|rrrrrrrrrrr}
\hline\hline
{\bf Observation} & Pointings & Beams/P  & FoV/P    & Res. &t$_\mathrm{int}$ &  BW  &  f & N$_\mathrm{ch}$& t$_\mathrm{samp}$    & Coh. & R$_{\rm d}$ \\
              &           &          & deg$^2$  & deg  & min & MHz  &  MHz &     & ms   &      &  TB/hr   \\
\hline
LPPS Survey   &  246      &  7       & 75       & 3.7  & 57    & 6.8  &  142 & 560 & 0.655   & I  & 0.09 \\
LOTAS Survey  &  206      &  19      & 3.9      & 0.5$^1$& 17  & 48   & 143 & 3904 & 1.3    &  C  & 0.8  \\
LOTAS Confirmation & $-$ & 61$-$217 & $-$       & <0.1  & 27   & 80  &  150  & 6576  & 0.49   & C & 12$-$42   \\
LOTAS Timing  &    $-$   &  1  & $-$  & $-$     & 15$-$30& 80  &  150  & 6576  & 1.3    & C       & 0.07 \\
\hline
\hline
\end{tabular}\\
\scriptsize{$^1$ With the 2011 beam-former, beams at higher declinations were spaced
more closely in right ascension \citepalias[\S 4.2.1 in][]{coen13}.}\vspace{2mm}
\caption{The different observational setups used in this work. Listed
  are the number of pointings over the survey; the number of beams per
  pointing Beams/P, the field of view per pointing FoV/P, the
  angular resolution (FWHM of each
  beam),
  integration time t$_\mathrm{int}$, bandwidth BW, central frequency
  f, number of channels N$_\mathrm{ch}$, sampling time
  t$_\mathrm{samp}$, whether beamforming was (I)ncoherent or (C)oherent, and the data rate R$_{\rm d}$.
\label{table-obs}}
\end{table*}

\subsection{LPPS Survey}\label{lpps-section}
The LPPS survey was conducted in 2010 December, using 
not-yet-calibrated Dutch LOFAR 
HBA stations \citep[cf.\ \S4.7 in][]{hwg+13}. Initial
observations used 13 core sub-stations and 4 remote stations, which 
 increased to 38 core and 6 remote stations at the end of the
 survey. As remote stations are twice as sensitive as core stations,
 the antenna gain $G$ for a certain LPPS configuration scales as
$G \propto (n_\mathrm{core} + 2 n_\mathrm{remote})/{\sqrt{n_\mathrm{core} + n_\mathrm{remote}}}$. 
As detailed in Ch.~3 of \citetalias{coen13}  
   these stations were combined 
 to form sets of 7 incoherent-sum station beams (Table \ref{table-obs}). 

\begin{figure*}[bt]
\centering
\includegraphics[width=\textwidth]{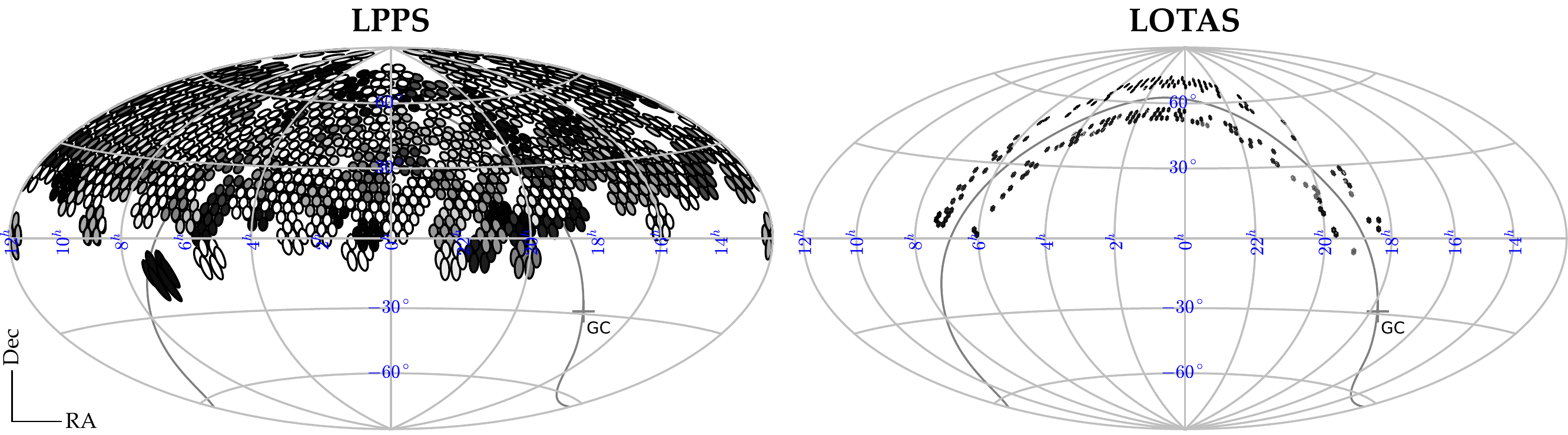}
\caption{The total sky coverage achieved by LPPS (left) and LOTAS
  (right). The Galactic plane and Center are shown with a grey line and
  cross, respectively. The grayscale of the individual beams shows the
  usable observation length, where white is 0\,min, and black the full
  57\,min for LPPS and 17\,min for LOTAS.  }
\label{lpps-plot-footprint}
\end{figure*}

Full-wave electro-magnetic simulations on HBA stations \citep{wijn12},
modeling the incoherent summation of the sets of stations used,
produced the FoV listed in Table \ref{table-obs}. The modeled LPPS beam
proved well-behaved and circular in the zenith, with a full width at
half maximum (FWHM) of
3.7\degr \, \citepalias[Fig.~3.1 in][]{coen13}.
The large FoV per pointing (Table \ref{table-obs}) allowed
the survey to cover 34\% of the celestial sphere
(Fig.\ \ref{lpps-plot-footprint}). Combined with the long dwell times,
effectively up to 1\,hr,
LPPS provided the equivalent of 9.7 minutes of 
\emph{all-sky} coverage.

\subsection{LOTAS Survey}
In May 2011 we carried out LOTAS, a tied-array survey of
$\sim$600\,deg$^2$.  By that time the 6 HBA stations on the central
Superterp 
shared a single clock signal\footnote{This signal is now distributed
  to all 24 core stations.} \citep{hwg+13}, which allowed for reliable
tied-array beamforming \citepalias[\S 4.2.1 in][]{coen13}. This resulted in
19-beam observations covering 3.9\,deg$^2$ per 17\,min pointing
(Fig.~\ref{fig-lotas-pointing-comparison}; Table \ref{table-obs}).
The expected tied-array beam shapes were validated through a series of
characterization observations \citep[Fig.~27 in][]{hwg+13}.
LOTAS sparsely surveyed two strips of sky $5\degr < |b| < 15\degr$, just off the Galactic
plane, to maximize the potential for new pulsar discoveries while avoiding
high scattering and background sky emission (Fig.\ \ref{lpps-plot-footprint}).

\subsection{LOTAS Confirmations and Timing}
\label{sect-time}
Confirmation observations on pulsar candidates from the LOTAS survey
were performed late 2012. 
The availability of 24 core stations, and ability
  to form many
tens of tied-array beams together enabled significantly higher sensitivity and
angular resolution (Table \ref{table-obs}). Confirmed candidates were
timed until early 2014 with a similar but single-beam setup (Table
\ref{table-obs}).


\section{Analysis}
 \label{sect:analyse}
Survey data were processed to detect periodic pulsar signals and
individual, dispersed bright radio bursts. This processing included
data quality checks. As the implementation of this pipeline and
quality control is detailed
in \citetalias{coen13}, only its essentials are summarized here. 

\subsection{Pipeline processing}
\label{section-pipeline}
Search processing was handled by a parallelized Python pipeline, based
on \texttt{PRESTO}\footnote{\url{https://github.com/scottransom/presto}}
\citep{2001PhDT.......123R}.  It automated the process of radio
frequency interference (RFI) excision
and dedispersion (with survey specific settings, described below). It
next handled periodicity searches including accounting for acceleration,
single-pulse searching, as well as sifting and folding of the best
candidates.

We first performed periodicity searches assuming no acceleration.
Binary motion, however, can cause the {\bfref intrinsically periodic signals} to
drift in the Fourier domain. Therefore we 
next performed an 'acceleration' search \citep{2002AJ....124.1788R} up to a maximum number of bins, or $z_{max}$, of 50.
For a pulsar with, e.g., a 40-ms period in a relativistic 
binary, LPPS is sensitive up to binary accelerations of
$51\,\mathrm{m\,s^{-2}}$ \citepalias{coen13}.  
We summed up to 16 harmonics, a technique where through recursive
stretching and summing, the power contained in all harmonics can be
recovered \citep{2001PhDT.......123R}; and included ``polishing'', an extra step to reject RFI
peaks acting as spurious harmonics.

\begin{figure*}
\centering
\includegraphics[width=0.66\textwidth]{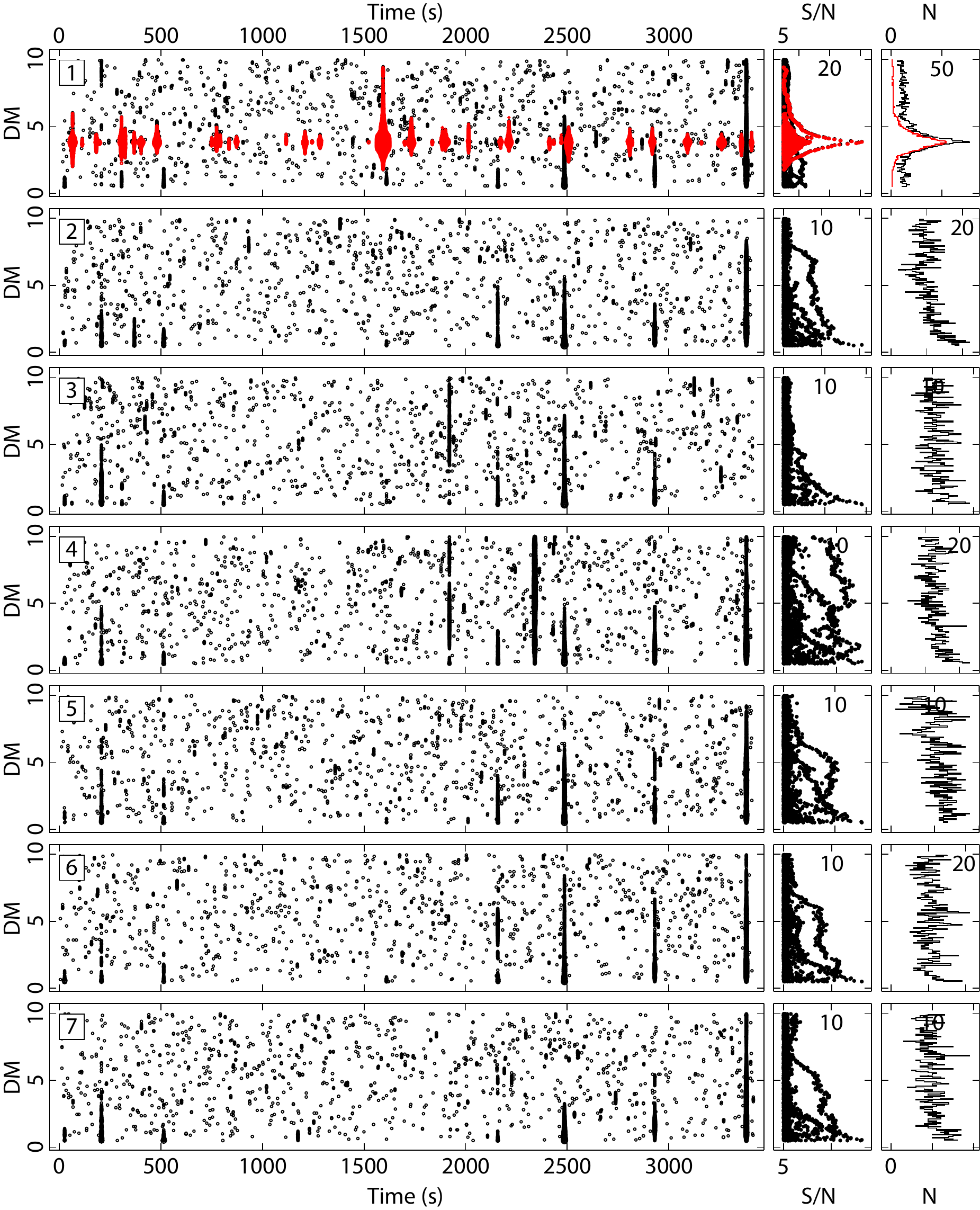}

\caption{A detection plot of PSR J0243+6257 (Beam 1, top row), as
  produced by our single-pulse post processing scripts. For each of
  the seven beams in this observation, the single-pulse detections are
  plotted between DMs of 0.5 cm$^{-3}$\,pc and 10 cm$^{-3}$\,pc (left
  panels).  The two right-most columns of panels show the events
  collapsed in time, in DM versus cumulative signal-to-noise ratio
  (S/N) and number of events (N), respectively.  The top panel of the
  plot shows PSR~J0243+6257 clearly detected whilst the other beams show
  no detection, only some RFI.  The pulsar is visible in the main
  top-row DM versus time panel as a series of individual pulses with a
  DM$\sim 3.9$\,pc cm$^{-3}$.  In
  post-processing, 
  our automated pulse grouping algorithm \citepalias[][]{coen13} has
  colored events that are judged to be of an astrophysical origin in
  red (the black points were automatically judged to be either
  statistical noise or RFI).}

\label{lpps-fig-j0243+65-7beams}
\end{figure*}

The dedispersed time series were inspected for single, dispersed
bursts by convolving the time series with a running box-car function
and inspecting the correlation coefficient.
We used box-cars 1--4, 6, 9, 14, 20 and 30 bins long, making the 
search sensitive to bursts between 0.655\,ms and 19.5\,ms at low DMs and, because
of down-sampling during dedispersion, to between 41.9\,ms and 1.26\,s at the very
highest DMs.
Data that were affected by RFI were discarded.  For the
remaining good-quality data, the \texttt{PRESTO} detections per individual
trial-DM were next associated or ``grouped'' \citepalias[\S 3.3.3 in][]{coen13} across all
DMs.  This freely
available\footnote{\url{https://github.com/tcoenen/singlepulse-search}}
single-pulse post-processing script looked for astrophysical signals
by demanding that each single-pulse detection (or ``group'') had a
minimum of 7 members, that it had a minimum signal-to-noise ratio of 8
and that at least 8 such pulses occurred at roughly the same peak
DM. Through trials, these numbers generally appeared to be the best heuristics
to distinguish astrophysical signals from man-made ones.
For each instance where these criteria were met, a plot like
Fig. \ref{lpps-fig-j0243+65-7beams} was created for inspection. Beyond
these, two-dimensional histograms of the number of detections in the
time-DM plane \citepalias[Fig.\ 3.5 in][]{coen13} were used to assess data quality and
detect bright pulsars.

\subsubsection{LPPS}\label{sect-lpps}
The LPPS data were reduced on the Hydra cluster at the University of
Manchester.  We dedispersed our data in 3487 trial-DMs up to 
$3000$\,pc cm$^{-3}$, and applied zero-DM filtering
\citep{2009MNRAS.395..410E}.  That technique removes all
broad-band signals from DM=0\,pc cm$^{-3}$, which are assumed to be
RFI. This proved useful in limiting the number of
spurious single-pulse detections without negatively impacting the
periodicity search results.
At the lowest DMs no downsampling in time was performed and the trial-DM
spacing was 0.05\,pc cm$^{-3}$.  At the very highest DMs the data were
downsampled 64 times, to match the increased intra-channel dispersive
smearing, and the trial-DMs are spaced in steps of 5\,pc
cm$^{-3}$.  Because of this heavy downsampling in time, the high-DM
part of the search added only a few percent to the total processing
time.

Acceleration candidates for all 7 beams were sorted by significance,
and duplicates in period and DM, across beams and
across $z_{max}$ values, were removed. To minimize spurious (RFI) detections, only candidates
with periods  $5\,\mathrm{ms} < P < 15\,\mathrm{s}$ and 
$DM > 2\,\mathrm{pc\;cm^{-3}}$ 
{\bfref were 
processed further.
The \texttt{prepfold} routine from  \texttt{PRESTO}
folded each candidate and optimized in period and period
derivative. Because this folding was done on the raw data, which includes frequency
information, the DM could also be optimized.
The resulting candidates were ranked.} Two neural nets 
\citep{2010MNRAS.407.2443E},  trained on the folds of LPPS pulsar
re-detections and RFI instances, helped prioritize candidates.

\subsubsection{LOTAS}\label{lotas-section-pipeline} 
LOTAS data were transferred from Groningen over a three-point
``bandwidth-on-demand'' 1$-$10\,Gbps
network,
to the grid storage cluster at
SURFsara\footnote{\url{https://www.surfsara.nl/nl/systems/grid/description}}
Amsterdam, and to Hydra.
The availability of a large grid compute cluster, operated by SURFsara
and part of the European grid infrastructure coordinated by
EGI\footnote{\url{http://www.egi.eu/}}, allowed processing of the full
survey data set to proceed relatively quickly. In all, the search ran
on 200 8-core servers for a little over a month, for a total of 1.3
million core hours used.

For LOTAS, the DM search covered $0 - 1000$\,pc\,cm$^{-3}$ with spacings
increasing from $0.02 - 0.30$\,pc\,cm$^{-3}$ for a total of either
16845 or 18100 DM trials \citepalias[cf. \S 4.2.3 in][]{coen13}. No zero-DM filtering
was used. All time series were searched at the full 1.3\,ms
resolution. 

LOTAS contained little RFI and no neural-net candidate pre-selection
was used. We manually inspected all acceleration candidates with
$\chi^2>2.0$. That reduced $\chi^2$, reported by \texttt{prepfold}, is
the result of fitting a straight line to the
profile -- {\bfref noisy profiles can be fit with such a straight line
  and will produce low $\chi^2$, while strong pulsars deviate and
  produce high $\chi^2$.} 
A $\chi^2>2.0$ corresponds to a
signal-to-noise ratio $\gtrsim$ 6. We also inspected 
those with $1.5 < \chi^2 <2.0$, DM\,$< 250$\,pc\,cm$^{-3}$ and
period derivative $\mathrm{\dot{P}} < 10^{-8} \mathrm{s}\,\mathrm{s}^{-1}$. 
{\bfref That choice of parameter space}
efficiently excludes RFI, which often shows large $\mathrm{\dot{P}}$.


\subsection{Telescope validation and data
  quality}\label{lpps-section-data-quality-control} 
The LPPS data were taken during LOFAR's early commissioning period and
data quality issues were expected.  Stations had not yet been
calibrated and contained faulty initial elements that generated RFI.
This affected the quality of the array beam, and a significant fraction
of the data was unusable\footnote{With system health monitoring now in
  place, the quality of current LOFAR data is {\it much}
  improved.}.

In the single-pulse search, RFI instances could sometimes overwhelm the 
\texttt{PRESTO} diagnostics. We developed a condensed version of the
single-pulse diagnostic plots, where the single-pulse candidates are
shown as a 2-dimensional, color-coded 
histogram on the time-DM plane \citepalias[Ch.\ 5 in][]{coen13}. 
Our removal of all 10-s blocks of data containing more than 500 single-pulse
candidates considerably cleaned the LPPS data set.

We obtained a first indication of the sensitivity of this uncalibrated
LOFAR  setup from the sample of  pulsars blindly detected in LPPS
(\S \ref{lpps-section-acceleration-results}). As detailed in \S3.4.4 in \citetalias{coen13}, we
extrapolated the ATNF catalog 400-MHz fluxes of this set to the LOFAR
band. From an ensemble comparison to our measured
signal-to-noise-ratio we found the maximum LPPS gain to be $G = 0.60
\pm 
0.13\;\mathrm{K/Jy}$. That is an appreciable fraction ($\sim$40\%) of
LOFAR's theoretical, calibrated sensitivity.

Compared to LPPS, the LOTAS data were already found to be \emph{much}
cleaner of both internally generated artifacts and external RFI.  This
was due to both the much more mature state of the deployed stations,
and the fact that the narrow, coherently formed beams are less
susceptible to RFI. The wider bandwidth of LOTAS, too, helped
differentiate interference -- which typically peaks at DM =
0\,pc\,cm$^{-3}$ -- from pulsar signals.

\section{Results and Discussion}\label{sec:results}

LPPS is a shallow survey that covered 34\% of the sky and detected
  65 known pulsars
(\S \ref{lpps-section-acceleration-results}). In \S
  \ref{lpps-section-popsyn}, we compare this yield with a simulation
  of 
the Galactic population. LPPS also
sets a stringent upper limit on the rate of low-frequency Fast Radio
Bursts (\S \ref{lpps-section-fast-transients-rate}). LOTAS, with its
higher sensitivity, detected the first two 
LOFAR pulsars (\S \ref{lotas-section-ps}).

\subsection{LPPS pulsar search}\label{lpps-section-acceleration-results}

The LPPS {blind periodicity} search yielded the re-detection of
54 pulsars. Their profiles are shown in
Figs. \ref{lpps-fig-profiles-1} and \ref{lpps-fig-profiles-2}. Folding
of {known pulsars}  within the
  survey footprint \citepalias[\S 4.2.4 in][]{coen13} led to the detection of
a further 9. The {single-pulse} search resulted in 20 pulsar
re-detections.
This means that
we were able to detect about one third of our periodicity search
detections in single-pulse searches of the same data.  This is in line
with predictions from \cite{2011A&A...530A..80S}.
Two pulsars, PSRs~B0154+61 and B0809+74, were
{only} detected through their individual pulses. 
In Table
\ref{lpps-table-redetections-ps} we list the detection parameters of all 
65 pulsars. 

The brightest {\bfref pulsar} in the Northern hemisphere, B0329+54, was detected in the side lobes
for many observations, from up to 109 degrees away from the pointing
center \citepalias[Fig.\ 3.14 in][]{coen13}.

Pulsar J2317+68 is a 813-ms pulsar at a DM of
71.158\,pc\,cm$^{-3}$, which we {independently} discovered.  This
source had only months earlier been discovered in the 350-MHz GBT
Northern Celestial Cap (GBNCC) survey \citep{slr+14}.

\begin{figure}
\centering 
\includegraphics[width=\columnwidth]{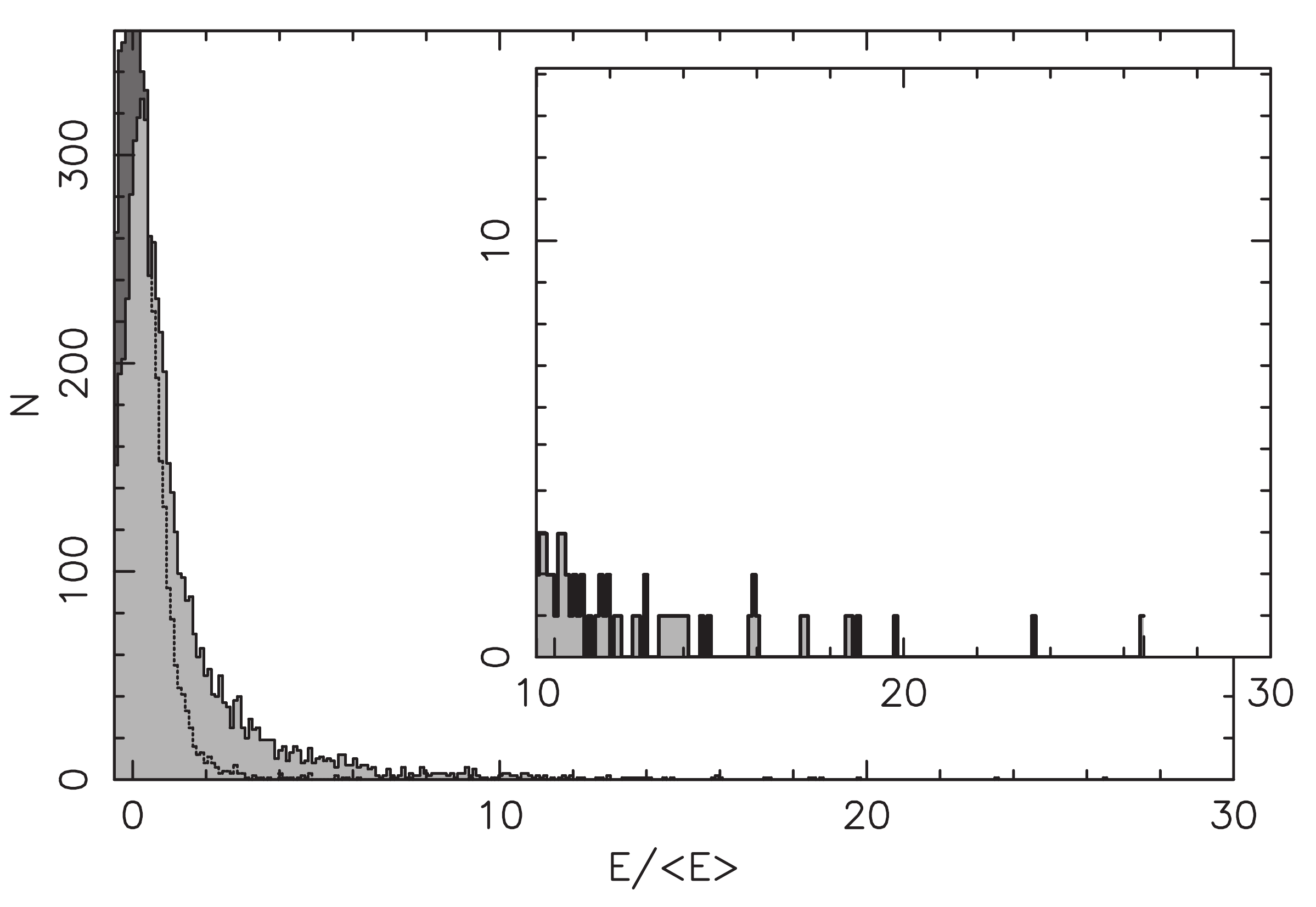}
\caption{Pulse energy histogram for PSR J0243+6257. Shown are the
  individual pulse energies for 6000 pulses, compared to their
  average. The dashed line, filled in dark gray, represents the
  off-pulse noise distribution. The light gray histogram is the energy
  present in the on-pulse region. That distribution peaks near zero
  and has significant overlap with the background histogram.  In those
  pulses no pulsar emission is detected.}\label{lpps-fig-e-hist}
\end{figure}

Pulsar  J0243+6257 was  discovered by the GBT350 Survey
\citep[][there known as J0240+62]{2008AIPC..983..613H}. It was
detected in both the LPPS periodicity and single pulse search. In many ways this source is a
prototype for the nearby, low-DM, perhaps intermittent, sources that
LOFAR is best equipped to discover. The
single-pulse discovery observation
(Fig.\ \ref{lpps-fig-j0243+65-7beams}) showed that PSR~J0243+6257 has
a broad pulse-energy distribution.  
In a 1-hr follow-up observation taken with the full LOFAR HBA core and
bandwidth, the large pulse-to-pulse intensity variations in PSR~J0243+6257
become even more clear. In Fig.\ \ref{lpps-fig-e-hist} we show that
some bright bursts outshine the average pulse by a factor 25. In most
of the well-studied pulsars this ratio is much lower \citep[e.g., a
  factor of only 2 in PSR~B0809+74; ][]{lkr+02}. It is only higher in
RRATs and PSR~B0656+14 \citep[where it is of order 100;][]{wwsr06}.
Using LOFAR's multi-beaming confirmation method (\S
\ref{sect-time}) this observation was also used to
determine the position to RAJ=02:42:35(3), DECJ=62:56:5(4).
With the long dwell
times that can be afforded by the huge FoV of LOFAR's incoherent
beam-forming mode, there are prospects for discovering other
intermittent but occasionally bright sources like PSR J0243+6257.

The detection of 54 pulsars in the blind search of the LPPS data
demonstrated that LOFAR is a capable pulsar search instrument. The
further detection of 9 pulsars by folding on their known ephemerides
means that our processing can still be improved {\bfref for better extraction of}
pulsars from spurious, RFI-related candidates. As LOFAR now routinely
produces better-calibrated, more sensitive data, the prospects for
achieving a significantly deeper all-sky survey using the same
incoherent beam-forming technique as LPPS is very good.

\subsubsection{Comparison to recent low-frequency or wide-field surveys}

Two features that make LPPS stand out from other recent pulsar survey
efforts (e.g.\ the Parkes multi-beam pulsar survey, \citealt{mlc+01};
the GBT driftscan survey, \citealt{bls+13}), are the very low
observing frequency and the large dwell time. Two surveys that each
\textit{did} share one of these characteristics are the second Cambridge
pulsar survey and the Allen Telescope Array (ATA) ``Fly's Eye''
experiment.

The second Cambridge survey at 81.5\,MHz
\citep[see][]{1998ApJ...509..785S} scanned the same northern sky as
LPPS and detected 20 pulsars. The blind periodicity search of LPPS
data did not detect 3 of those (B0809+74, B0943+10, B1133+16) because
their LPPS pointings were corrupted.  B1642$-$03 fell
outside our survey area. 

Notable features of LPPS are its large FoV, and 57-min dwell time. It
shares these features with the ``Fly's Eye'' experiment carried out
with the ATA \citep{wbb+09} at a frequency of 1.4\,GHz. Its
single-pulse searches were the \mbox{L-Band} 
equivalent of those in LPPS. Down to a
signal-to-noise ratio of 8 no astronomical single pulses were 
detected beyond those from PSR~B0329+54 \citep{sbf+12}.

\subsection{Modeling the pulsar population that underlies LPPS}
\label{lpps-section-popsyn}

On completion LPPS was the
deepest large-area survey in the 100\,MHz regime to date. 
The 65 pulsar
  detections constitute a low-frequency sample that is adequate in
  size to serve as input for a pulsar population model.

\begin{figure*}
\centering
\includegraphics[width=0.8\textwidth]{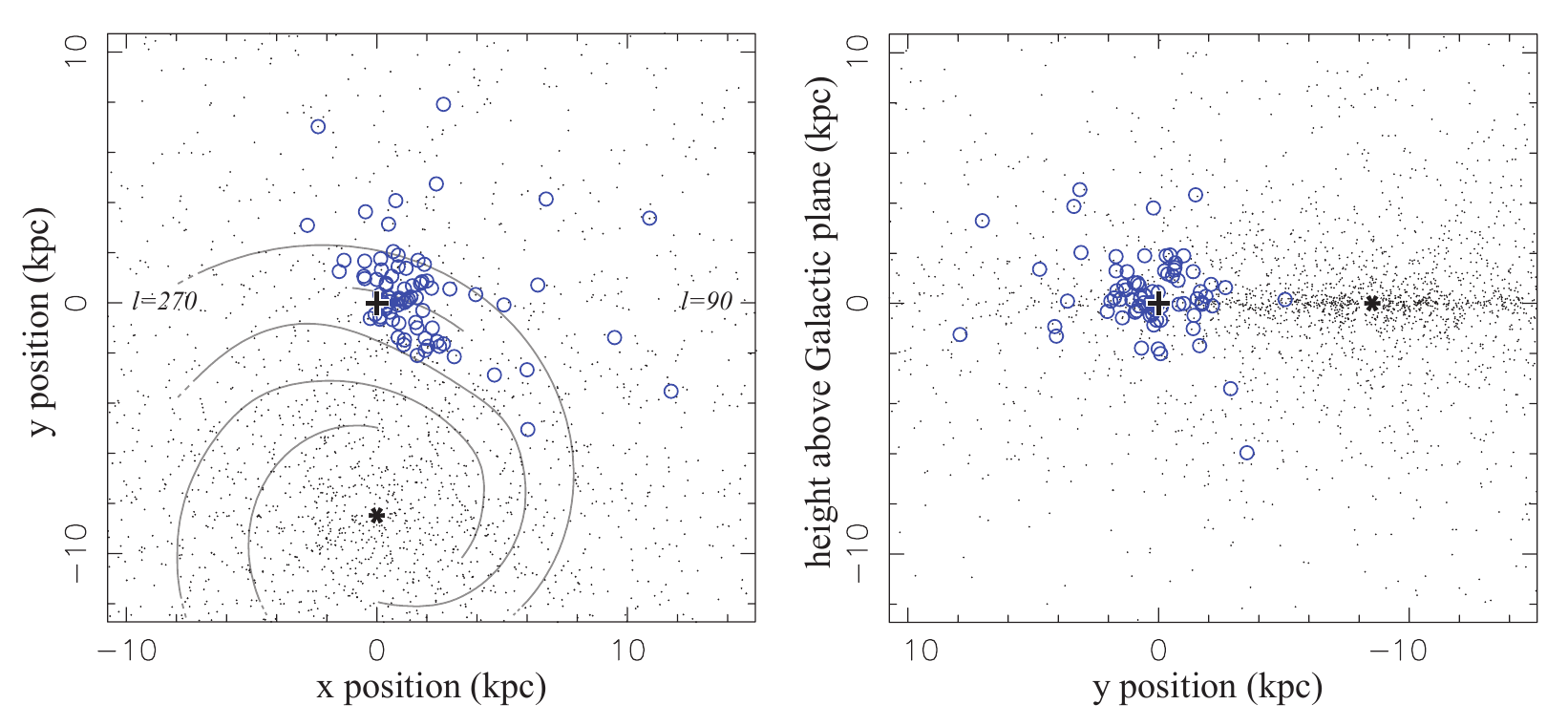}
\caption{ The modeled pulsars in our Galaxy. Pulsars that are still
  emitting at the current time, and are beamed toward us, are marked
  with dots (only 10\% of these shown). Simulated pulsars that are detected in
  LPPS are marked with blue open circles. The Earth is marked with a `+', the
  Galactic center with a `*'. 
  On the left a projection onto the plane of the Galaxy is plotted,
  including spiral arms, with the Galactic centre at y=$-$8.5\,kpc from
  the Earth. The right hand panel shows the projection of the detections
  onto the vertical plane through the Galactic centre and the Earth.
}
\label{lpps-fig-xyz}
\end{figure*}

In modeling LPPS we used the dynamic, evolving model approach developed
in \citet{hbwv97a}. We populated the modeled Galaxy with a population
of pulsars that in \citet{ls10} {\bfref best reproduced} the survey
results of 6 large surveys. We implemented a survey model of LPPS that 
takes into account the total footprint on the sky,
including overlapping regions (Fig.\ \ref{lpps-plot-footprint}); the
distribution of usable integration times; the sensitivity variation
between pointings using different sets of stations; and the gain
determined in \S \ref{lpps-section-data-quality-control}. Based on the
cumulative signal-to-noise-ratio values determined for our detected sample
(Table \ref{lpps-table-redetections-ps}), any simulated pulsar 
producing a S/N over 15 was labeled detected. That S/N value is higher
than the common value of 8$-$10 \citep[cf.][]{ls10} to account for the deleterious effect that RFI had on
  blindly identifying pulsar signals in LPPS.

In our simulation, 2.7 million pulsars formed throughout the
Galaxy. Of these, 50,000 were above the death line at the present day
and beamed toward Earth; 9,000 are in our survey FoV. For 1,200 of
these, the scatter and dispersion broadening exceeds their rotational
period, making them undetectable. Of the remaining 8,000 pulsars, 80
are bright enough to be detected in LPPS
(Fig. \ref{lpps-fig-xyz}). Running the simulation over the error range
on the derived gain of 0.60$\pm$0.13$\;\mathrm{K/Jy}$ produces a LPPS
detected sample containing 80$\pm$20 simulated pulsars. That is in reasonable
agreement with the actual number of blind detections of 54.  Some of the
difference between these numbers could be in either the modeled survey
(e.g., remaining incomplete understanding of the incoherent-addition
of these commissioning era data) or the modeled population (e.g., the
low-frequency spectrum turn-over behaving differently than
simulated). Overall, these simulations confirm that our best pulsar
population models \citep{ls10} can accurately predict low-frequency
surveys.

\subsection{LPPS limit on the rate of fast radio bursts}
\label{lpps-section-fast-transients-rate}

The detection of a bright, highly dispersed radio burst of apparent
extra-galactic origin was first reported by
\citet[][]{2007Sci...318..777L} and further detections have since been
reported by \citet{kkl+11}, \citet{2013Sci...341...53T} and
\citet{2014arXiv1404.2934S}.
As the LPPS survey provides both long
dwell times and large FoV, the survey data can be used to either
detect or limit FRBs at low radio frequencies.

We searched the single-pulse data (\S \ref{section-pipeline}) down to
a signal-to-noise ratio of 10 at DMs between 2--3000 cm$^{-3}\;$pc.
This is a much larger DM than predicted for any typical line-of-sight
through our Galaxy away from the Galactic center
\citep{2002astro.ph..7156C}. Signals with such high DMs may also be
highly scattered at low frequencies. Although there is an observed
relation between DM and scattering delay in the Galaxy
\citep{2004ApJ...605..759B}, this relation has more than an
order-of-magnitude scatter and any single line-of-sight may deviate
significantly from the average relation.  Furthermore, it is unlikely
that the ionized inter-galactic medium is distributed in a similar way
to the Galactic inter-stellar medium.  Such highly dispersed signals
may thus continue to be detectable at LOFAR frequencies
\citep{2013IAUS..291..492S,2013ApJ...776..125M,2013MNRAS.436..371H,2013MNRAS.436L...5L},
making a detection possible and a limit useful.

We visually inspected all pulses that crossed this threshold of S/N
$> 10$.  All were associated with either a known pulsar or RFI (as
evidenced by detections across multiple beams, or across multiple
trial DMs with no peak in S/N) which was particularly present at low
DMs. Thus, no FRBs were detected in LPPS.

For bursts that are not affected by intra-channel dispersion smearing, the
resulting fast-transient limit then depends on the telescope 
sensitivity and the sky coverage in both time and area.  We define
the sky coverage as being out to the FWHM of each beam and use the
effective time coverage determined in \S
\ref{lpps-section-data-quality-control}.

We use the peak gain derived from the LPPS pulsar re-detections (\S
\ref{lpps-section-data-quality-control}) to calculate the
sensitivity.  For each LPPS observation this was adjusted for the
number of used (sub-)stations and the zenith angle, and was multiplied by
0.73 to produce the average over the FoV.  Use of the gain
  average, not its minimum at FWHM, is common in
determining survey sensitivity \citep[e.g.\ ][]{ebsb01}.
Finally, we derived a flux
limit for each pointing using a $T_{\mathrm{sys}} + T_{\mathrm{sky}}$
of 500\,K.  We removed from consideration any beam with a flux limit
above 200\,Jy, which was likely caused by RFI and/or calibration
issues. The LPPS average flux limit is then S$_\mathrm{min}$ = 107\,Jy
\citepalias[Fig.\ 3.15 in][]{coen13}. 

That limit is valid when the observed burst, after
dedispersion, falls within a single 0.66\,ms sample. For bursts that
are wider, either intrinsically or by intra-channel or step-size
dispersion smearing, the burst flux is spread over a width $w$. The
fluence $F=Sw$ is preserved when a pulse is smeared out
\citep[e.g. ][]{2013Sci...341...53T} but the sensitivity of our
single-pulse searches over boxcars of up to 1.26\,s (\S
\ref{section-pipeline}) decreases by the square root of this width $w$.

We thus find a rate limit $R$ on FRBs that were detectable by LPPS of
\begin{equation}
   \hspace{10mm} 
   R_\mathrm{FRB} (S > 107 \sqrt{\frac{w}{0.66\,\mathrm{ms}}}\,{\rm Jy}) < 
   1.5\times 10^{2} \,{\rm sky^{-1} day^{-1}} .
\end{equation}

\begin{figure}
\centering
\includegraphics[width=0.9\columnwidth]{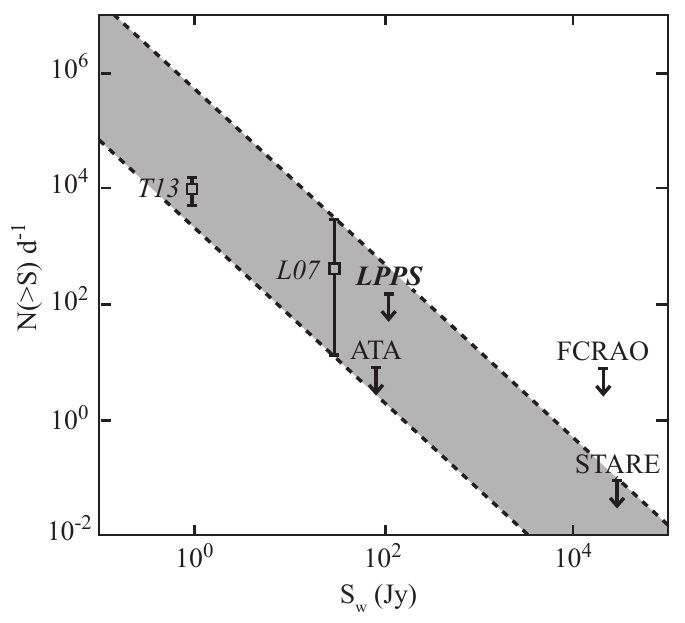}

\caption{The limits on transient occurrence per day $N$ versus
  width-adjusted minimum flux
  $S_w = S/\sqrt{\frac{w}{0.66\,\mathrm{ms}}}$, on a logarithmic scale, comparing previous fast-transient
  surveys \citep{koj+08} with LPPS.  The  burst rates reported by
  \citet[][L07]{2007Sci...318..777L}
  and  \citet[][T13]{2013Sci...341...53T} are plotted with their
  errors. The
  area between the dashed lines represents a
  $N(>$$S)\propto S^{-3/2}$ prediction from a homogeneous,
  stationary population of objects.  The LPPS limit,  which assumes a
  flat spectral index for the purposes of comparing with surveys at
  other frequencies, is seen near the center. 
}

\label{lpps-fig-n_vs_s}
\end{figure}

In Fig.\ \ref{lpps-fig-n_vs_s} we compare our rate to the
extrapolated occurrence based on the detection of the first 
FRB \citep{2007Sci...318..777L}, to the 
 FRB rate reported in \citet{2013Sci...341...53T} and to limits set by other surveys.  These
other surveys were all performed at higher frequencies,
270--1400\,MHz. The low observing frequency used in LPPS means that
any steep spectrum sources would appear much brighter, and
consequently that a flux limit obtained with LOFAR would be much more
constraining when extrapolated to higher observing frequencies. The
original detection of the Lorimer burst had an apparently steep
spectrum \citep{2007Sci...318..777L}, while one of the 
 bursts reported in \citet{2013Sci...341...53T} instead showed 
100-MHz-wide, bright bands. Given our limited
knowledge at this time, we simply assume a flat spectral index.

We can convert this celestial rate to a volumetric event frequency.
We start from the assumptions that FRB emission is intrinsically
shorter than our sampling time of 0.655\,ms, has a luminosity at
1.3\,GHz of 1\,Jy\,Gpc$^2$ \citep{2013Sci...341...53T}, and follows a
power-law scaling of the fluence $F(\nu) \propto\ \nu^\alpha$.
We assume our intra-channel dispersion smearing dominates
over scatter smearing.

For each LPPS beam we determine at which distance the decreasing
flux of such an FRB falls under the minimum detectable flux for the
increasingly dispersion-smeared pulse.
As the distance to a simulated 
 FRBs increases, we calculate its flux $S$, which falls per the 
 inverse-square law based on the co-moving distance. At each step in
 redshift $z$, we calculate the expected DM by adding the maximum Galactic
 contribution in this direction 
\citep{2002astro.ph..7156C}, a host contribution, and an intergalactic
matter (IGM) component. 
We assume an intrinsic host DM of 100\,pc\,cm$^{-3}$ and a reduction
of the effective time
smearing with redshift as 1/(z+1).
From Fig.~1 in \citet{2003ApJ...598L..79I} we
estimate DM$_\mathrm{IGM} \simeq 1100\,z $, for $z <
4$. We assume the intrinsic dispersion in the burst is negligible.
From the combined effective DM we determine the number of time bins $n$ the burst is smeared
over. After also taking into account the average expected mismatch
with the closest box-car length (\S \ref{section-pipeline}) this results in an increase of
the previously determined minimum
detectable flux S$_\mathrm{min}$
by a factor $\sqrt{1.125 n}$. The distance at which
this $S_\mathrm{min}$ > $S$ determines the volume this beam has
searched. This is multiplied by the pointing integration time.  Given
our non-detection, the reciprocal of the resulting 
summed flux-limited volume over all such beams then produces a 
rate upper limit of
\begin{equation}
   \hspace{12mm} 
   \Phi_\mathrm{FRB} < 2.5 \times 10^{5} ~ 
   \left(\frac{142}{1300}\right)^{-1.3 (\alpha+2)} \,
   \mathrm{Gpc}^{-3}\,\mathrm{yr}^{-1} .
\end{equation}

Here the dependence of the deleterious dispersion smearing on frequency 
causes the power-law index to deviate from the $-1.5(\alpha+2)$ expected for a purely
flux-limited cosmological population. 
For $\alpha=-2$ our upper limit 
of $2.5 \times 10^{5} \,{\rm Gpc^{-3}\,yr^{-1}}$ is higher than, and thus consistent with,
the \citeauthor{2013Sci...341...53T} FRB rate at 1.3\,GHz of 
$2.4\times 10^{4}$ Gpc$^{-3}$\,yr$^{-1}$ as derived in
\citet{2014arXiv1402.4766K}. For shallower spectral indexes our rate upper
limit increases further and becomes less constraining.

The limits we find
are in line with
those derived from other surveys.
In ongoing LOFAR transient searches,
we are continuing to improve on this limit and better constrain the
spectra and scattering properties of such bursts.  These searches will
either soon detect such signals or show that the higher-frequency
($\sim$1.4\,GHz) window is ideal for their detection.

\subsection{LOTAS pulsar search}\label{lotas-section-ps} 

\begin{figure}
\includegraphics[width=\columnwidth]{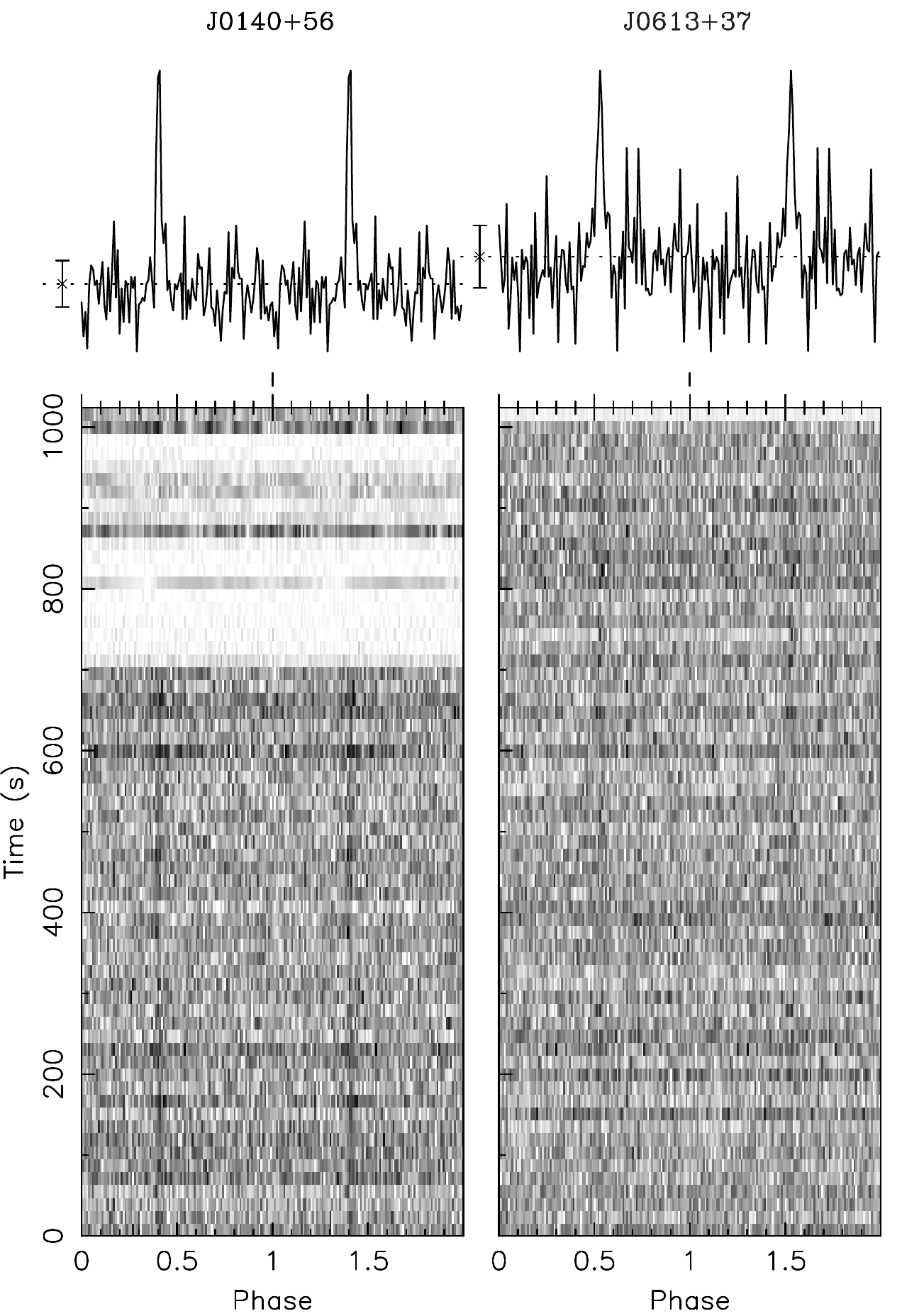}
\caption{Discovery plots of PSRs J0140+5622 (left) and  J0613+3731
  (right). 
  The bottom panels shows
  the signal strength as a function of time and rotational phase. Some
  RFI is masked, showing 
  up as whiteouts. The top panels shows the folded profiles, repeated
  twice for clarity. \label{lotas-fig-discovery-folds} }
\end{figure}

The LOTAS periodicity search resulted in the detection of 23 pulsars,
including LOFAR's first two pulsar discoveries: PSRs~J0140+5622 and
J0613+3731 (Fig.\ \ref{lotas-fig-discovery-folds}).
The parameters of all
detections are listed in Table \ref{lotas-table-detections}. Of the 21
redetections, 17 were available in the ATNF pulsar catalog.  A further
four, J0216+52, J0338+66, J0358+42 and J2243+69, were independent
discoveries that had only recently been found in the ongoing GBNCC survey
\citep{slr+14}.  

To judge the completeness of our blind search, we directly folded at
the periods of all known pulsars within a 5-degree radius of the LOTAS
pointing centers. Inspecting these folds yielded the re-detection of a
further 6 pulsars. These are marked ``d" in Table
\ref{lotas-table-detections}. 

The LOTAS single-pulse search output suffered from temporal
misalignment between subsequent DM trials, which prevented automatic
separation of astrophysical pulses from RFI. By manually inspecting the
condensed plots produced by the single-pulse search (\S 
\ref{lpps-section-data-quality-control}) we identified 8 bright
pulsars.  In Table \ref{lotas-table-detections} we mark these ``s" and
list the S/N of the brightest pulse. We then checked for all known
pulsars within a 7\,degrees radius of the LOTAS pointing, but we did not detect any.

For the two discoveries,
PSRs J0140+5622 and J0613+3731,
confirmation and follow-up observations 
(\S \ref{sect-time}) were combined to determine the rotational ephemerides.
These data sets 
span two years.
This timing analysis was performed using \texttt{PSRCHIVE}
\citep[][]{2004PASA...21..302H,sdo12}  and
\texttt{TEMPO2}
\citep[][]{2006MNRAS.369..655H} and followed standard techniques
\citepalias[detailed in][]{coen13}.
The resulting timing solutions are given in Table \ref{table-lotas-timing}.

When the most recent of these timing observations were carried out in
2014 January, the LOFAR sensitivity was well described and
characterized, allowing for the flux measurements also listed in  Table
\ref{table-lotas-timing}. As detailed in \citet{khs+14}, these take
into account e.g., the actual number of healthy dipoles used, the
empirical scaling of sensitivity versus number of coherently added
stations, and the exact fraction of masking due to RFI.

\subsubsection{PSR J0140+5622}
LOFAR's first pulsar discovery, PSR J0140+5622, has a spin period $P =
1.8$\,s and DM = 101.8\,pc\,cm$^{-3}$.  Its discovery profile is shown
in Fig.~\ref{lotas-fig-discovery-folds}. Our timing campaign
allowed us to derive a period derivative $\mathrm{\dot{P}}$ of
$7.9\,\times\,10^{-14}$\,s\,s$^{-1}$.  The values derived for the
characteristic age $\tau_c$ and the surface magnetic field
$B_{\mathrm{surf}}$ are respectively 3.5\,$\times\,10^{5}$\,yr and
1.2\,$\times\,10^{13}$\,G. 
This surface magnetic field is at the high end of the
known population \citepalias[Fig.\ 4.14 in][]{coen13}.
The best-fit timing model for PSR J0140+5622 is presented in 
Table \ref{table-lotas-timing}. The estimated distance to
this pulsar is 3.8\,kpc, based on the NE2001 model
\citep{2002astro.ph..7156C}.

\subsubsection{PSR J0613+3731}
\begin{figure}
\centering
\includegraphics[width=\columnwidth]{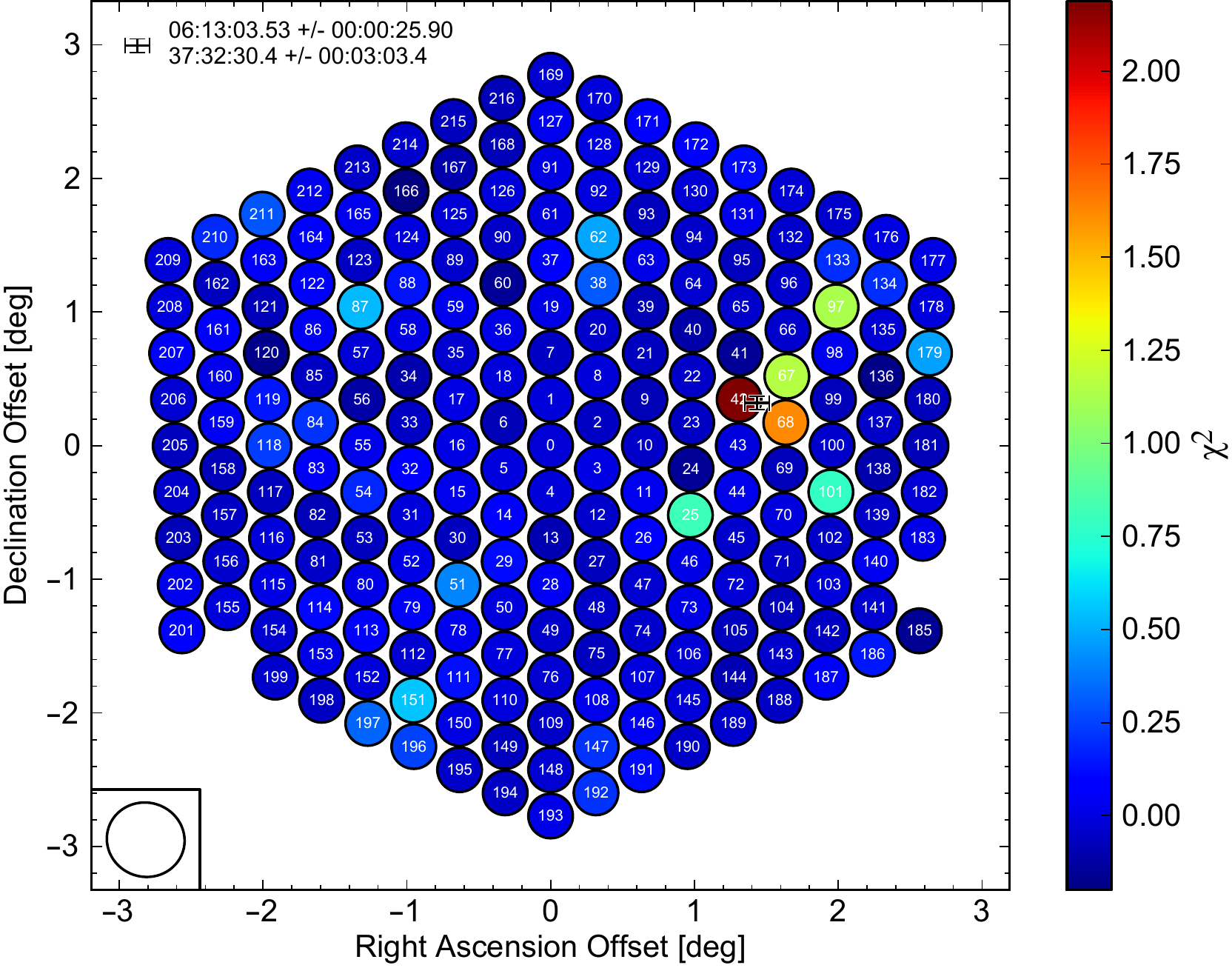}

\caption{The confirmation observation of PSR J0613+3731, establishing
  it had been discovered in a side-lobe of the original survey
  observation. The initially derived position was between beams 15 and
  30 in this figure. Two of the 217 specified beams did not process
  due to a cluster hardware issue. Each beam is color coded according
  to the reduced chi-squared of the average pulse profile in that
  beam, compared to a straight line (cf. \S
  \ref{lotas-section-pipeline}). The actual size of the individual 
  tied-array beams is shown bottom left. The cross shows the pulsar's
  best-fit position.}

\label{fig-lotas-localization-215beams-J0607+37}
\end{figure}

The second LOTAS pulsar discovery is PSR J0613+3731
(Fig. \ref{lotas-fig-discovery-folds}), with $P = 0.62$\,s and DM =
19.0\,pc\,cm$^{-3}$.  While this pulsar was quickly confirmed in a
multi-beam observation, 
its observed position appeared to vary, and it was only visible at
the bottom of the observing band. Side-lobe detections of bright known
pulsars showed similar behavior. As the LOTAS fractional bandwidth is large,
the FWHM of a tied-array beam changes by
$\sim$40\% over the band,
and the size and position of the side lobes follow.
A further localization observation
performed as part of the timing campaign for PSR J0613+3731, this time
using 215 beams spread over an area much larger than the 19-beam
footprint of a LOTAS survey observation, established the fact that PSR
J0613+3731 was indeed initially discovered in a side-lobe
(Fig. \ref{fig-lotas-localization-215beams-J0607+37}).  The timing
campaign along with the initial discovery and confirmation
observations allowed us to derive values for $\mathrm{\dot{P}}$,
$B_{\mathrm{surf}}$ and a $\tau_c$, which are respectively
$3.2\,\times\,10^{-15}\,$s\,s$^{-1}$, $1.4\,\times\,10^{12}$\,G and
$3.0\,\times\,10^{6}$\,yr. Table \ref{table-lotas-timing}
contains the precise values for these derived quantities and also
contains information about the timing solutions. The DM-derived
distance to PSR~J0613+3731 is 0.64\,kpc.

\begin{table*}
\centering
\begin{tabular}{p{80mm}ll}
\hline\hline
\multicolumn{3}{c}{Measured and derived quantities} \\
\hline
Pulsar name\dotfill & J0140+5622                                  & J0613+3732 \\                     
Right ascension (J2000), $\alpha$\dotfill &  01:39:38.561(19)             &  06:13:12.149(11) \\              
Declination (J2000), $\delta$\dotfill & +56:21:36.82949(1)                & +37:31:38.30520(1) \\             
                      
Galactic Latitude, $l$ (degrees)  \dotfill & 129.6090   & 175.3357   \\
Galactic Longitude, $b$ (degrees) \dotfill & $-$5.8853    & 9.2388    \\

Pulse frequency, $\nu$ (s$^{-1}$)\dotfill & 0.56327117338(3)      & 1.61499182415(16) \\              
Frequency derivative, $\dot{\nu}$ (s$^{-2}$)\dotfill & $-$2.50951(8)$\times 10^{-14}$  & $-$8.463(5)$\times 10^{-15}$ \\   
Dispersion measure, $DM$ (cm$^{-3}$pc)\dotfill & 101.842(32)      & 18.990(12) \\                        
Reference epoch (MJD) \dotfill & 56000                           & 56000 \\                          
MJD range\dotfill & 55693.4$-$56691.7                             & 55693.6$-$56665.9 \\              
Number of TOAs\dotfill & 25                                       & 74 \\                             
Rms timing residual ($\mu s$)\dotfill & 1151.7                    & 996.9 \\                          
Flux density at 0.15\,GHz, $S_{0.15}$ (mJy)\dotfill & 4(1)   & 13(1)   \\
Flux density at 0.35\,GHz, $S_{0.35}$ (mJy)\dotfill & 0.7(3) & 1.6(8)  \\
Flux density at 1.4\,GHz,  $S_{1.4}$  (mJy)\dotfill & 0.02(1)& 0.13(6) \\ 
Spectral index 0.15$-$0.35\,GHz,  ${\alpha}_{0.15-0.35}$ \dotfill & $-$2.1(1.0) & $-$2.5(1.3) \\ 
Spectral index 0.35$-$1.4\,GHz,  ${\alpha}_{0.35-1.4}$   \dotfill & $-$2.6(1.7) & $-$1.8(1.2) \\ 

Fractional pulse width $w_{50}$  \dotfill & 0.024 & 0.017 \\ 
Characteristic age, $\log_{10}$($\tau_{\mathrm{c}}$ (yr)) \dotfill & 5.55               & 6.48 \\                           
Surface magnetic field, $\log_{10}$($B_{\mathrm{surf}}$ (G))   \dotfill & 13.08  & 12.16 \\                          
\hline                                                                                                
\end{tabular}%
\vspace{1mm}%
\caption{Measured and derived quantities for the two newly discovered pulsars. Figures in parentheses are  the nominal 1$\sigma$ 
\texttt{TEMPO2} uncertainties in the least-significant digits quoted.
For the fluxes, errors are estimated to be 50\%.
The pulse widths are quoted as fraction of the pulsar period $P$.
 \label{table-lotas-timing}}
\end{table*}

\subsubsection{Follow-up}\label{lotas-section-discussion}
High-frequency confirmation attempts were carried out at 1532\,MHz
with the Lovell telescope at Jodrell Bank Observatory. 
Both pulsars were ultimately detected.
Pulsar J0613+3731 was seen easily, although in follow-up timing it
showed considerable flux variation possibly caused by
scintillation, explaining why it was  missed in previous 1.4\,GHz
surveys.
Pulsar J0613+3731 thus showcases the benefit of a
low-frequency survey, where the observed bandwidth greatly exceeds the
scintillation bandwidth.
From a 4-hr integrated profile we 
estimate an L-Band mean flux density of 0.13(6)\,mJy, using the
radiometer equation \citepalias[Eq.~3.3 in][]{coen13}
to scale to the off-pulse noise. 
Pulsar J0140+5622 is much fainter at 1.4\,GHz. Only after a total of
6\,hrs of integration this pulsar was detected at a S/N of 10 --
clearly illustrating the importance of
low-frequency surveys for finding steep-spectrum sources.
The resulting average flux density is listed in Table
\ref{table-lotas-timing}.

Next, both sources were weakly detected in the expected archive
pointings from GBNCC \citep{slr+14}, by folding at their now-known
periods.  The resulting 350\,MHz mean flux densities, derived using
the radiometer equation, are listed in Table \ref{table-lotas-timing}.
Between each of the derived fluxes at 150, 350 and 1400\,MHz we
derived the spectral index $\alpha$ from a fit to $S_\nu \propto
 \nu^\alpha$, where $S_\nu$ is the flux density at observing frequency
 $\nu$. Table  \ref{table-lotas-timing} shows these are steeper than
 $-$2 at the low-frequency end, for both pulsars.

\section{Conclusions and future
  work}\label{overall-section-conclusion}

LPPS re-detected 54 pulsars in the
periodicity search, and 20, mostly overlapping, in single pulse
searches. A further 9 pulsars were retrieved from the data by folding
on known ephemerides (Table \ref{lpps-table-redetections-ps}). 
That detected sample agrees well with the outcome of a simulation of
the Galactic pulsar population and the LOFAR telescope response.
Making
use of the large LPPS footprint, in both sky coverage and time, we
derived a limit on the occurrence of FRBs of width $w$ and flux
density 
$S > 107 \sqrt{\frac{w}{0.66\,\mathrm{ms}}}\,{\rm Jy}$
of no more than
$1.5\times10^{2}$ day$^{-1}$ sky$^{-1}$ at 142\,MHz.

LOTAS produced the first two pulsars ever
discovered with LOFAR.
Periodicity searching also found 21 known
pulsars. Of these, 8 were also re-detected through single-pulse
searches. A further 6 pulsars were detected by folding data on known 
ephemerides. 

LPPS and LOTAS were surveys in their own right -- discovering new
pulsars and setting the first low-frequency FRB limit.  They were also
critical learning steps towards a proper LOFAR pulsar survey.  
The full LOFAR pulsar search, the LOFAR
Tied-Array All-Sky (LOTAAS, \emph{``with double A''}) survey,
includes aspects of both LPPS and LOTAS; it
simultaneously creates both large incoherently formed beams and many
tied-array beams using the 6 Superterp stations
(Fig.~\ref{fig-lotas-pointing-comparison}, right panel). This setup allows the best
of both worlds; the large FoV afforded through incoherent beam-forming
and its concomitant sensitivity to rare bright bursts, and the raw
sensitivity afforded by coherent beam-forming.

LOTAAS is using many more tied-array beams than LOTAS, as well as a
much more complicated pointing strategy that takes better advantage of
LOFAR's flexible multi-beaming capabilities.  Each survey pointing is
comprised of three sub-array pointings (i.e. three beams generated at
station level).  An incoherent array beam is generated for each of
these sub-array pointings, and together these cover $\sim$60\,square
degrees of sky at a sensitivity roughly twice that of LPPS.  Within
the FoV of each incoherent beam we also form a Nyquist-sampled,
hexagonal grid of 61 tied-array beams.  Together, this set of $3
\times 61$ tied-array beams covers a survey area of $\sim$12\,square
degrees at a sensitivity roughly twice that of LOTAS and the GBNCC.
Through 3 interlinked overlapping survey passes our sensitivity for
intermittent or transient sources is greatly improved.  Processing for
LOTAAS has begun at the University of Manchester and on the new Dutch
national supercomputer
Cartesius\footnote{\url{https://www.surfsara.nl/nl/systems/cartesius}}.
Extrapolating from the pilot surveys, LOTAAS could discover at least
200 new pulsars over the whole northern hemisphere.

\subsection{The Square Kilometre Array}
Looking beyond LOFAR, the two pulsar discoveries reported here are the
first ever using a sparse digital aperture array.  We are confident
that these discoveries validate our approach to pulsar surveys, an
important point as the upcoming Phase I of the Square Kilometre Array
(SKA) will feature a LOFAR-like low-frequency aperture array of $\sim$250,000 dipole antennas \citep{2010iska.meetE..18G}, more than an
order-of-magnitude more collecting area than LOFAR. These will be
beam-formed into stations in a LOFAR-like way. The LPPS and LOTAS
surveys act as a proving ground for such flexible digital beam-forming
capabilities.  The techniques pioneered by LOFAR and presented here
will thus be important as precursors to the eventual pulsar surveys
with the SKA.

\begin{acknowledgements}
We thank J.~Green and P.~Abeyratne for help with LPPS processing. 
This work was supported by grants from the Netherlands Research School for
Astronomy (NOVA3-NW3-2.3.1) and by the European Commission
(FP7-PEOPLE-2007-4-3-IRG \#224838) to JVL; 
an NWO Veni Fellowship to JWTH; and
Agence
Nationale de la Recherche grant ANR-09-JCJC-0001-01 to CF.
 LOFAR is designed
and constructed by ASTRON, and is operated by the International LOFAR
Telescope (ILT) foundation.  LOTAS ``bandwidth-on-demand'' was
provided by SURFnet.  LOTAS processing was carried out on the Dutch
grid infrastructure supported by the BiGGrid project.

\end{acknowledgements}

\newpage

\begin{appendix}
\onecolumn
\section{LPPS detections and profiles}
\label{sec:positions:lpps}
\begin{longtable}{|l|l|l|l|l|l|l|l|}

\caption{\label{lpps-table-redetections-ps}All pulsars detected in the
  LPPS data. Sources marked with an asterisk were used to derive an
  estimate for the LPPS gain (see \S
  \ref{lpps-section-data-quality-control}). The column headed by an M
  (method) lists how the pulsar was detected --- through a periodicity
  search (p), through a direct fold using a known ephemeris (d), or
  through single-pulse searches (s). The next column, headed by {\it
    $\alpha$}, lists the distance to the beam center in degrees at
  which the pulsar was detected (based on the ATNF pulsar database
  positions).  The {\it DM} and {\it period} columns list the values
  as reported by \texttt{PRESTO} (without errors or digit
  significance) after folding the data on the known ephemeris 
  using 256 bins. The next two columns report the {\it peak} and the
  {\it cumulative} signal-to-noise ratio derived from the pulse
  profiles. For profiles where the
  off-pulse baseline is not flat (because of RFI), the signal-to-noise
  ratios are left blank. The last column shows the peak signal-to-noise ratio of the
  brightest pulse found in the single-pulse search as reported
  by \texttt{PRESTO} (not corrected for zero-DM filtering).}\\

\hline
Pulsar & M & $\alpha$ ($\deg$)&DM (pc\,cm$^{-3}$)&Period (ms)&S/N$_\mathrm{p}$&S/N$_\mathrm{cum}$&S/N$_\mathrm{s}$\\
\hline
\hline
\endfirsthead
\caption{Continued}\\
\hline
Pulsar&M&$\alpha$ ($\deg$)&DM (pc\,cm$^{-3}$)&Period (ms)&S/N$_\mathrm{p}$&S/N$_\mathrm{cum}$&S/N$_\mathrm{s}$\\
\hline
\hline
\endhead
\hline
\endfoot
B0045+33	&p	&1.53	&39.940	&1217.0		&14  	&-	&-\\
B0105+65	&d	&1.99	&30.460	&1283.6		&9	&93	&-\\ 
B0136+57$^*$	&p	&2.57	&73.779	& 272.4		&7	&74	&-\\
B0138+59$^*$	&p,s	&2.67	&34.797	&1222.9		&9	&29	&7\\
B0144+59	&d	&1.61	&40.111	& 196.3		&7	&64	&-\\ 
B0154+61	&s	&1.37	&30.00	&-		&-	&-    &9\\ 
{J0243+6257}	&p,s	&-	&3.903	&591.7		&9	&36	&31\\
B0329+54$^*$	&p,s	&2.44	&26.833	&714.5		&606	&2137	&-\\ 
B0355+54$^*$	&p,s	&1.10	&57.142	&156.3		&32	&352	&7\\
B0402+61	&p	&1.65	&65.303	&594.5		&8	& 82	&-\\ 
B0450+55$^*$	&p	&2.20	&14.495	&340.7		&32	&237	&-\\
{J0540+3207}	&p	&-	&62.371	&524.2		&9	&-    &-\\
B0523+11	&p	&0.85	&79.345	&345.4		& 7	&149	&-\\
B0611+22$^*$	&p	&1.53	&96.910	&334.9		&10	&208	&-\\
B0626+24	&d	&2.03	&84.195	&476.6		& 4	& 35	&-\\ 
B0655+64$^*$	&p	&0.59	&8.771	&195.6		&81	&550	&-\\
B0809+74	&s	&2.66	&5.75	&-		&-	&-    &14\\ 
B0823+26$^*$	&p,s	&1.66	&19.454	& 530.6  	&120	&560	&22\\
B0834+06$^*$	&p	&1.50	&12.889	&1273.7  	&226	&347	&-\\
B0917+63	&p,s	&0.50	&13.158	&1567.9  	& 20	&243	&11\\
B0919+06$^*$	&p	&2.09	&27.271	& 430.6  	& 92	&618	&-\\
B0950+08$^*$	&p,s	&2.23	&2.958	& 253.0  	& 60	&658	&21\\
B1112+50	&p,s	&1.68	&9.195	&1656.4  	&-  	&-    &24\\ 
B1237+25$^*$	&p,s	&1.92	&9.242	&1382.4  	& 27	& 46	&31\\
B1322+83	&p	&1.61	&31.312	& 670.0  	&  7	&-	&-\\
B1508+55$^*$	&p	&2.10	&19.613	& 739.6  	&303	&1451	&-\\
B1530+27	&p,s	&1.41	&14.698	&1124.8  	&-	&-	&10\\
B1541+09$^*$	&p	&1.67	&35.240	& 748.4  	&26	&247	&-\\
B1604$-$00	&p	&2.70	&10.682	& 421.8  	&-	&-	&-\\ 
B1633+24	&p	&1.73	&24.320	& 490.5  	&-	&-	&-\\ 
J1758+3030	&p	&1.33	&34.900	& 947.2  	&-	&-	&-\\ 
B1811+40$^*$	&p	&0.94	&41.487	& 931.0  	&30	&309	&-\\
B1821+05$^*$	&p	&1.65	&66.775	& 752.9  	&25	&192	&-\\
B1839+09	&p	&0.75	&49.107	& 381.3  	&12	&-    &-\\
B1839+56	&p	&2.58	&26.298	&1652.8  	&- 	&-    &-\\ 
B1842+14$^*$	&p	&1.55	&41.510	& 375.4  	&29	&147	&-\\
B1905+39$^*$	&p,s	&1.55	&30.960	&1235.7  	&11	&38	&12\\
B1918+26	&p	&0.39	&27.620	& 785.5  	&18	&136	&-\\ 
B1919+21	&p,s	&2.39	&12.455	&1337.3  	&-	&-	&32\\
B1929+10	&p	&2.14	&3.180	& 226.5  	&-	&-	&-\\
B1949+14	&d	&1.44	&31.460	& 275.0  	&  4	& 39	&-\\
B1951+32	&d	&0.54	&45.006	&  39.5  	&  8	&223	&-\\
B1953+50$^*$	&p,s	&1.29	&31.974	& 518.9  	& 28	& 96	&12\\
B2016+28$^*$	&p,s	&0.72	&14.172	& 557.9  	&124	&156	&14\\
B2020+28$^*$	&p	&0.43	&24.640	& 343.4  	& 50	&314	&-\\
B2021+51$^*$	&p,s	&0.71	&22.648	& 529.1  	& 18	&182	&30\\
B2022+50$^*$	&p	&1.64	&33.021	& 372.6  	&  9	& 28	&-\\
J2043+2740$^*$	&p	&1.42	&21.000	&  96.1  	& 22	&276	&-\\
B2110+27	&p	&2.95	&25.113	&1202.8  	& 36	&199	&-\\
B2111+46$^*$	&p,s	&1.66	&141.260&1014.6  	& 24	&483	&12\\
J2139+2242	&p	&0.79	&44.100	&1083.5  	& 11	&114	&-\\
B2154+40	&p	&2.09	&70.857	&1525.2		&-	&-	&-\\
B2217+47$^*$	&p,s	&1.80	&43.519	& 538.4		&447  	&1976	&12\\
B2224+65$^*$	&p	&2.86	&36.079	& 682.5		&7	& 55	&-\\
B2227+61	&d	&0.71	&124.614& 443.0		&5	&178	&-\\
B2255+58	&p	&1.26	&151.082& 368.2		&-	&-    &-\\ 
J2302+6028	&d	&2.20	&156.700&1206.4		&6	& 76	&-\\
B2303+30	&p	&1.14	&49.544	&1575.8		&-	&-    &-\\ 
B2306+55$^*$	&p	&2.75	&46.538	& 475.0		&5	& 13	&-\\
B2310+42$^*$	&p	&1.78	&17.276	& 349.4		&25	&215	&-\\
B2315+21	&p	&2.11	&20.906	&1444.6		&23	&-    &-\\
{J2317+68}	&p	&-	&71.156	& 813.3		&7	& 35	&-\\
B2319+60	&d	&1.75	&94.591	&2256.4		&2	&-    &-\\
B2334+61$^*$	&p,s	&2.14	&58.410	& 495.3		&11	& 90	&13\\
B2351+61	&d	&0.15	&94.662	& 944.7		&10 	&-	&-\\ 
\end{longtable}

\newpage

\begin{figure}[h]
\centering
{\includegraphics[trim=1cm 3cm 1cm 1cm, clip=true, width=0.8\textwidth]{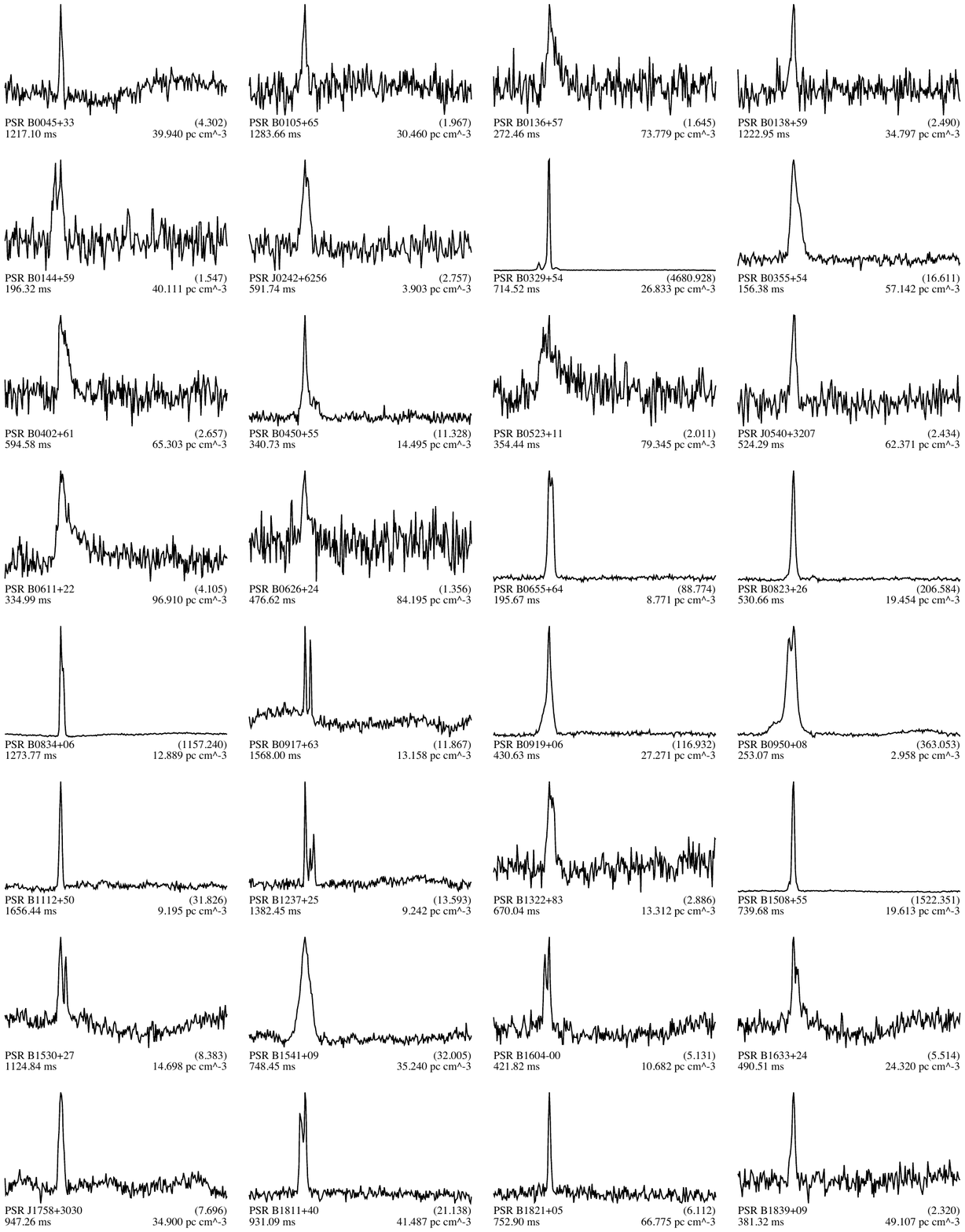}}

\caption{LPPS pulsar profiles. For each pulsar, the catalog name,
  detected pulse period, 
  detected DM, and reduced chi-squared significance (in parentheses)
  from the automated search fold are given.  One full rotational cycle
  {\bfref is shown}.  Some of the brightest pulsars were detected multiple
  times; only the highest signal-to-noise detection is shown here.
  Several of the pulse profiles show scattering tails, e.g. PSRs
  B0523+11, B0611+22, B2111+46, making them excellent sources for
  studies of the interstellar medium.  In many cases the off-pulse
  baseline is not flat due to RFI that could not be completely
  excised.}

\label{lpps-fig-profiles-1}
\end{figure}

\begin{figure*}[h]
\centering
\includegraphics[trim=1cm 3cm 1cm 1cm, clip=true, width=0.8\textwidth]{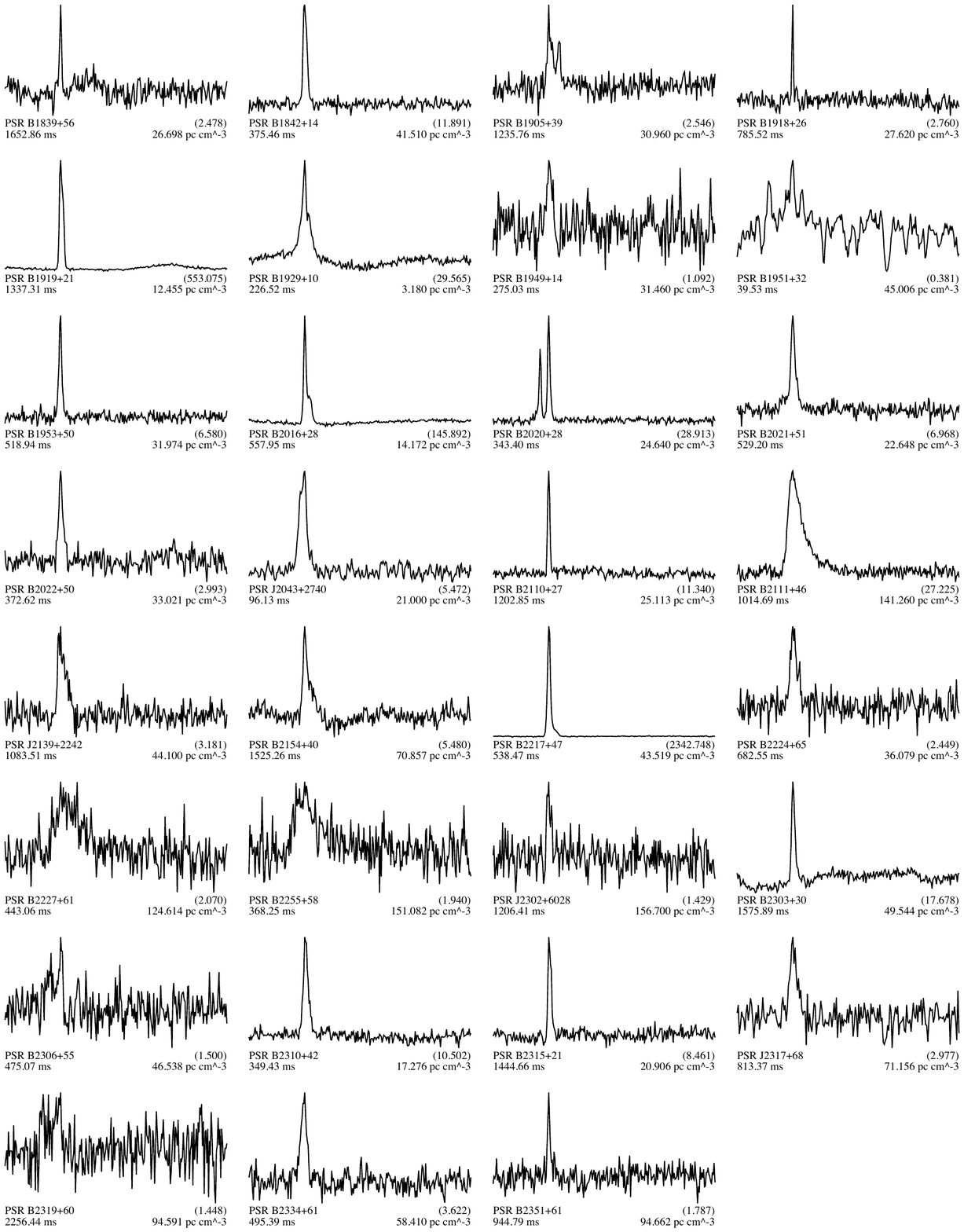}
\caption{LPPS pulsar profiles -- continued.}
\label{lpps-fig-profiles-2}
\end{figure*}

\newpage
~\\
\newpage

\section{LOTAS detections and profiles}
\label{sec:positions:lotas}
\begin{table*}[h]
\centering
{\small
\begin{tabular}{|l|l|l|l|l|l|l|l|}
\hline
Pulsar&M&$\alpha$ (deg)&DM (pc\,cm$^{-3}$)&Period (ms)&S/N$_{\mathrm{p}}$&S/N$_{\mathrm{cum}}$&S/N$_{\mathrm{s}}$\\
\hline
\bf{J0140+5622}  &p	&0.16	&101.637&1775.4	&9	&40	&-\\ 
\bf{J0613+3731}  &p	&1.52	&19.106	& 619.1	&5	&42	&-\\ 
\hline
\hline
B0037+56  	&d	&0.19	&92.480	&1118.1	&10	&85	&-\\ 
B0105+65  	&d	&3.32	&30.681	&1283.6	& 5	&22	&-\\ 
B0136+57  	&p	&3.00	&73.894	& 272.4	& 9	&44	&-\\ 
\bf{J0216+52} &p	&-	&22.030	&  24.5 &20	&94	&-\\ 
B0329+54  	&p,s	&4.09	&26.785	& 714.5	&19	&73	&10\\
\bf{J0338+66}  	&p	&-	&66.577	&1761.9	& 6	&- 	&-\\ 
\bf{J0358+42}  	&p	&-	&46.308	& 226.4	& 5	&56	&-\\ 
B0402+61  	&p	&0.82	&65.360	& 594.5	& 8	&40	&-\\ 
B0450+55  	&p,s	&0.10	&14.600	& 340.7	&138	&512  	&35\\
B0525+21  	&p,s	&2.80	&51.133	&3745.5	&9	&-	&17\\
J0611+30  	&p	&1.95	&45.320	&1412.2	&3	&-	&-\\ 
B0609+37  	&d	&1.37	&27.084	& 297.9 &3	&10 	&-\\ 
B0626+24  	&d	&2.80	&83.523	& 476.6	&4	&-	&-\\ 
B1821+05  	&p,s	&0.87	&66.754	& 752.9	&40	&70	&8\\ 
J1822+0705	&p	&0.20	&62.282	&1362.8	&8	&39	&-\\ 
B1839+09  	&d	&3.20	&49.146	& 381.2 &4	&22	&-\\ 
B1911$-$04  	&p	&2.40	&89.380	& 825.9	&37	&132	&-\\ 
B1918+26  	&p,s	&0.07	&27.673	& 785.5	&51	&100	&10\\
B1919+21  	&p	&3.74	&12.394	&1337.2	&19	&26	&-\\ 
J1942+01  	&p	&0.09	&52.267	& 217.3	&5	&43	&-\\ 
B1953+50  	&p	&1.29	&32.036	& 518.9	&11	&6	&-\\ 
J2007+0910	&d	&3.08	&47.291	& 458.6	&5	&38	&-\\ 
B2154+40  	&p,s	&0.88	&71.281	&1525.2	&23	&68	&9\\ 
B2217+47  	&p,s	&1.45	&43.463	& 538.4	&400	&913	&43\\
B2224+65  	&p	&2.02	&36.333	& 682.5	&23	&112	&-\\ 
B2241+69  	&p,s	&0.17	&40.820	&1664.4	&31	&43	&10\\
\bf{J2243+69}  	&p	&-	&67.732	& 855.4	&5	&35 	&-\\ 
\hline
\end{tabular}
\vspace{2mm}
}
\caption{A list of all LOTAS pulsar detections. Pulsar names in bold
  are newly discovered pulsars, the first two rows contain unique
  LOFAR discoveries and the other pulsars with bold names signify
  \emph{independent} discoveries --- i.e. unpublished sources recently
  found by competing surveys such as the GBNCC \citep{slr+14}. The $M$ column
  contains the method through which the pulsar was detected: (p) in
  the blind periodicity search, (d) through folding on a known
  ephemeris, and (s) by inspecting condensed single-pulse plots. The
  $\alpha$ column gives the 
  distance to the center of the nearest tied-array beam. These distances
  are based on the ATNF catalog value. For the
  LOTAS discoveries the timing position
  was used.  
  The
  DMs and periods quoted are taken from the \texttt{PRESTO}
  diagnostics of the folds; these
  values were optimized during the folding process in order to
  maximize the signal-to-noise ratio. There is no error or digit
  significance information for these.
  The columns S/N$_\mathrm{p}$ and S/N$_{\mathrm{cum}}$ quote
  respectively the peak and cumulative signal-to-noise ratios if they
  could be derived automatically. The signal-to-noise ratios were
  derived from profiles with 100 bins.  The S/N$_{\mathrm{s}}$ column contains the
  signal-to-noise ratio of the brightest single pulse detected,
  possibly in another beam than the periodicity search detection.}
\label{lotas-table-detections}
\end{table*}

\newpage 

\begin{figure*}[h]
\centering
\includegraphics[width=0.8\textwidth]{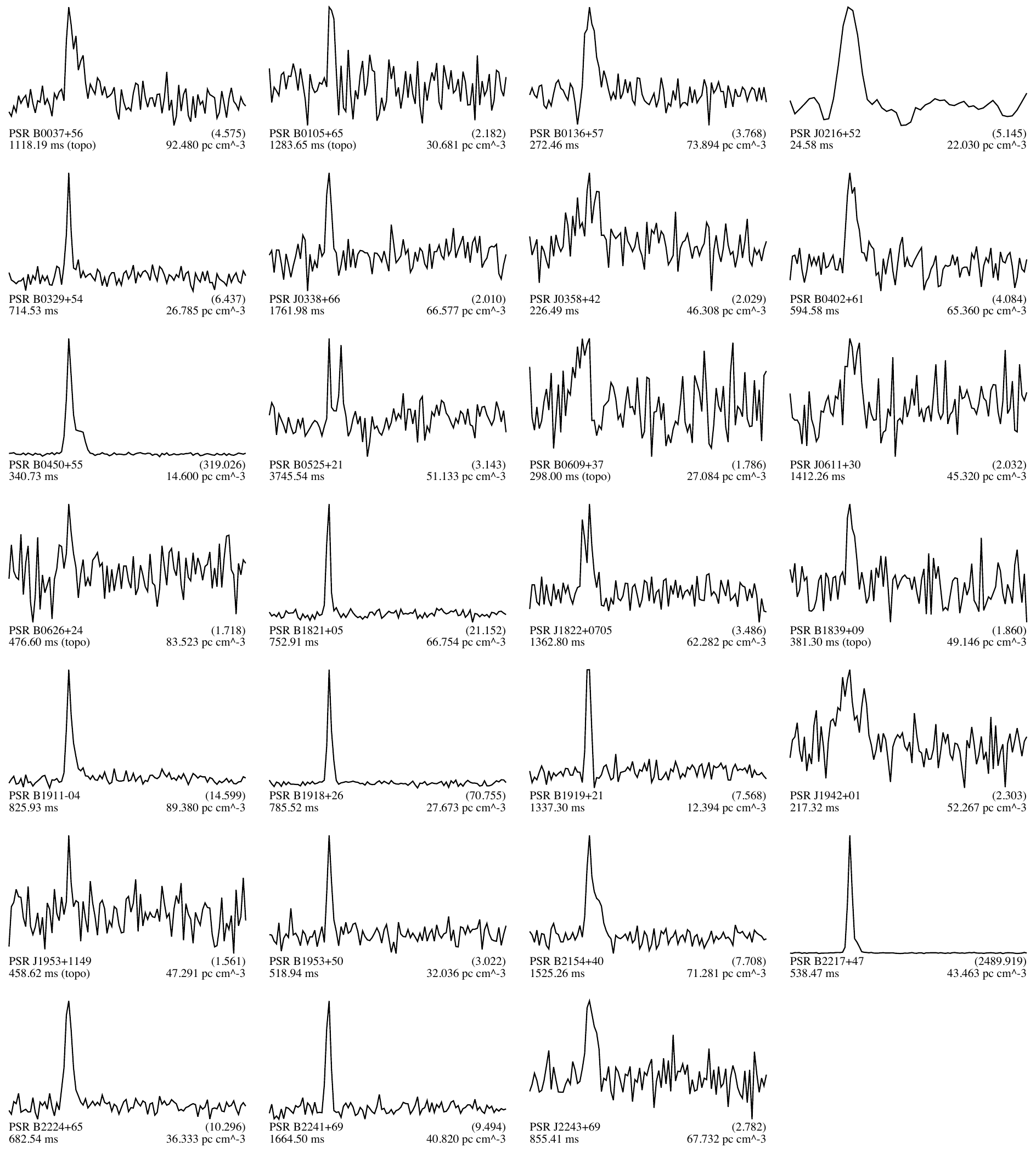}
\caption{Average pulse profiles for pulsars re-detected in LOTAS. Both
  pulsars found in the blind periodicity search and pulsars
  re-detected by folding on previously known ephemerides are
  shown. For each pulsar the name, period, reduced chi-squared (in
  brackets) and DM are quoted. The reduced chi-squared is a proxy for
  the signal-to-noise of the pulse profile and is reported by \texttt{PRESTO}.
  Periods marked with ``(topo)'' are topocentric values as produced by
  the direct folding feature of the pipeline (\S \ref{lotas-section-ps}).}
\label{lotas-fig-profiles}
\end{figure*}

\end{appendix}

\end{document}